\documentclass[sigconf]{acmart}
\usepackage{color,soul}
\usepackage[section]{placeins}

\usepackage{xr}
\usepackage{stfloats}

\newcommand{\revised}[1]{\textcolor{black}{#1}}

\AtBeginDocument{%
  \providecommand\BibTeX{{%
    \normalfont B\kern-0.5em{\scshape i\kern-0.25em b}\kern-0.8em\TeX}}}

\setcopyright{acmcopyright}

\copyrightyear{2023} 
\acmYear{2023} 
\setcopyright{acmlicensed}\acmConference[CHI '23]{Proceedings of the 2023 CHI Conference on Human Factors in Computing Systems}{April 23--28, 2023}{Hamburg, Germany}
\acmBooktitle{Proceedings of the 2023 CHI Conference on Human Factors in Computing Systems (CHI '23), April 23--28, 2023, Hamburg, Germany}
\acmPrice{15.00}
\acmDOI{10.1145/3544548.3581330}
\acmISBN{978-1-4503-9421-5/23/04}

\begin{document}

\title[When do Data Visualizations Persuade?]{When do data visualizations persuade? The impact of prior attitudes on \revised{learning about correlations} from \revised{scatterplot} visualizations}


\author{Douglas Markant}
\email{dmarkant@uncc.edu}
\orcid{0000-0003-0568-2648}
\authornotemark[1]
\affiliation{%
  \institution{University of North Carolina at Charlotte}
   \city{Charlotte}
   \country{USA}
}

\author{Milad Rogha}
\email{mrogha@uncc.edu}
\orcid{0000-0002-1464-2157}
\affiliation{%
  \institution{University of North Carolina at Charlotte}
  \city{Charlotte}
  \country{USA}
}

\author{Alireza Karduni}
\email{akarduni@uncc.edu}
\orcid{0000-0001-9719-7513}
\affiliation{%
  \institution{University of North Carolina at Charlotte\\ IDEO}
   \city{Chicago}
   \country{USA}
}

\author{Ryan Wesslen}
\email{ryan@explosion.ai}
\orcid{0000-0001-9638-8078}
\affiliation{%
 \institution{Explosion}
   \city{Charlotte}
   \country{USA}
}

\author{Wenwen Dou}
\email{wdou1@uncc.edu}
\orcid{0000-0003-0319-9484}
\affiliation{%
  \institution{University of North Carolina at Charlotte}
   \city{Charlotte}
   \country{USA}
}


\begin{abstract}
\revised{Data visualizations are vital to scientific communication on critical issues such as public health, climate change, and socioeconomic policy. They are often designed not just to inform,} but to \textit{persuade} people to make consequential decisions (e.g., \revised{to get vaccinated}). Are such visualizations persuasive, especially when audiences have beliefs and attitudes that the data contradict? In this paper we examine the impact of existing attitudes (e.g., positive or negative attitudes toward COVID-19 vaccination) on \revised{changes in beliefs about statistical correlations when viewing scatterplot visualizations with different representations of statistical uncertainty}. We find that strong prior attitudes are associated with smaller belief changes when presented with data that contradicts existing views, and that visual uncertainty representations may amplify this effect. Finally, \revised{even when participants' beliefs about correlations shifted} their attitudes remained unchanged, highlighting the need for further research on \revised{whether data visualizations can drive longer-term changes in views and behavior}.
\end{abstract}

\ccsdesc[500]{Human-centered computing~Visualization}

\keywords{data visualization, persuasion, attitudes, uncertainty communication, science communication}


\maketitle

\section{Introduction}

Data visualizations are an increasingly central part of science communication, fueled by the growing accessibility of public data, advances in the usability of visualization software, and the rise of data journalism~\cite{weber_data_2018}. 
Visualizations are used to inform the public about pressing social issues such as public health (e.g., COVID-19 trends), democractic backsliding (e.g., polarization, gerrymandering), and climate change (e.g., weather anomalies, natural disasters).
However, while these efforts have increased the accessibility of scientific evidence, their impact on the public's knowledge and attitudes is often unclear, with some evidence that people struggle to engage with and learn from such visualizations~\cite{matheus_data_2020}.

This discrepancy between the accessibility and impact of public data was on stark display during the COVID-19 pandemic that began in 2020.
During this period the public struggled to navigate uncertainty over the health risks and effectiveness of public health measures related to the COVID-19 virus~\cite{noar_miscommunicating_2020}.
Yet while scientific consensus on these matters gradually solidified, in the United States the early politicization of public health measures contributed to entrenched divides in attitudes toward COVID-19 preventive behaviors~\cite{hegland_partisan_2022}.
This polarization also likely influenced how people consumed information about the pandemic. 
Recent evidence suggests that people interpret scientific evidence about COVID-19 in a way that is biased by preexisting beliefs or attitudes~\cite{hutmacher_role_2022}, view evidence-based recommendations as untrustworthy or partisan~\cite{ternullo_im_2022}, and are susceptible to misinformation that undermines official public health messaging~\cite{van_der_linden_misinformation_2022}.
These factors may have also played a role in how people engaged with ``crisis  visualizations'' that tracked the course of the pandemic~\cite{zhang_mapping_2021}, with recent work suggesting that political leaning affects how people interpret visualizations related to COVID risks~\cite{ericson_political_2022}.
However, despite their prominence during the pandemic, relatively little is known about how public opinion and behavior were influenced by interactions with such data visualizations.

The uncertain impact of COVID-19 crisis visualizations exemplifies a broader need to understand the \textit{persuasive power} of data visualizations when they challenge people's existing views~\cite{Pandey_TVCG_2014}. 
 \revised{Statistical evidence is commonly used to persuade and change behavior, and in certain cases is perceived as more persuasive than other forms of evidence like narratives and anecdotes~ \cite{allen1997comparing, allen2000testing}. As journalists and science communicators move toward using visualizations to present statistical evidence, can we expect such visualizations to hold similar persuasive power?}
While existing research has examined a wide range of \revised{perceptual and cognitive} factors that concern how people perceive and understand visualizations~\cite{franconeriScienceVisualData2021}, less work has considered how existing attitudes influence how people 
\revised{interact with, learn from, and change attitudes in light of} 
data visualizations.
As a result, relatively little is known about the kinds of visualizations that are best able to persuade people to change their minds and behavior.

Our goal in this study was to investigate how existing attitudes affect how people update their beliefs about \revised{statistical} relationships depicted in data visualizations.
We build on prior work by Karduni \textit{et al.}~\cite{karduni_bayesian_2021} which investigated changes in beliefs about bivariate correlations when interacting with scatterplots with different visual representations of statistical uncertainty.
In the present study we examined how attitudes about two topics (COVID-19 vaccination and labor union membership) affect the extent to which people change their beliefs about relevant empirical phenomena.
In addition, we investigated the impact of visual representations of statistical uncertainty on belief change.
The use of uncertainty representations in science communication and data visualization has been the subject of debate, with some concern that they limit the clarity or persuasiveness of the intended message~\cite{kelp_vaccinate_2022,Hullman8805422}.

Our findings reveal that strong preexisting attitudes about a topic \revised{were associated with} smaller changes in beliefs after viewing \revised{scatterplots with data that are inconsistent with prior beliefs}.
In addition, this effect was more pronounced for \revised{scatterplots that included visual encodings of statistical uncertainty (animated or static confidence intervals), suggesting that visual representations of uncertainty may reduce the persuasiveness of such visualizations when the data conflicts with existing attitudes.}
We also found that, while participants adjusted their beliefs about specific empirical relationships, there was little evidence for systematic changes in attitudes regarding either topic after interacting with the visualizations, consistent with prior literature from social psychology on the difficulty of changing well-established attitudes~\cite{skitka2015psychological}. 
These findings provide novel insights into how existing attitudes, which can be rooted in personal and cultural factors unrelated to the presented data, shape how people interpret data visualizations, and call for more research on \revised{how to design visualizations that can persuade people with views that are incongruent with the data.}

\section{Background and Related Work}

\subsection{Beliefs, attitudes, and persuasion}

Psychological theories of persuasion broadly distinguish between changes in \textbf{beliefs} and changes in \textbf{attitudes}~\cite{Albarracin:2005}, a crucial distinction for understanding the role of data visualizations in changing people's minds. 
A \textbf{belief} is a person's agreement with the truth of a claim, which can be represented as a proposition about the relationship between entities or variables (e.g., \textit{``Vaccines cause autism''}).
Individuals might differ in their belief in a particular proposition, but beliefs about observable phenomena should change as new empirical evidence comes to light, just as a scientist might update a theory in response to anomalous data.
From this perspective, the proximate goal of a data visualization is to \textit{inform}: To change beliefs about a specific empirical relationship (e.g., by showing that the incidence of autism is unrelated to childhood vaccination rates).
Of course, people might fail to update their beliefs for many reasons, including because they generate interpretations of the evidence that are more consistent with an existing belief or
question the validity or trustworthiness of its source~\cite{chinnRoleAnomalousData1993}.

Even when beliefs do change after seeing new evidence, that may not be accompanied by broader shifts in attitudes.
Whereas beliefs concern specific claims, \textbf{attitudes} are overall evaluations of entities or issues \revised{in a positive or negative light} (e.g., being pro-vaccination vs. anti-vaccination).
People may draw on a knowledge base of beliefs to evaluate or justify their attitudes (e.g., \textit{President Biden is honest} is one of many interrelated beliefs that might affect the global attitude expressed in presidential approval ratings), but those beliefs are not the only source of justification for attitudes which may be intertwined with other aspects of social identity, emotion, and broader worldviews~ \cite{Albarracin:2005,bryant_mass_2002,eagly1993psychology, zanna8p}. 
\revised{Attitudes are also held with varying degrees of certainty, with past work suggesting that attitudes on some issues are vague and become more well-defined with repeated expression or increasing experience in a domain}~\cite{Petrocelli2007}.
Strongly held attitudes can be difficult to change because of their relationships to personal values and identity~\cite{petty1986communication} and can persist even when people change their beliefs about \revised{related claims}~\cite{bryant_mass_2002}.
For instance, Nyhan \textit{et al.} found that correcting specific misconceptions about vaccination (i.e., by presenting evidence that vaccines are not associated with increased rates of autism) had no effect on overall attitudes toward vaccination or behavioral intentions~\cite{nyhanEffectiveMessagesVaccine2014}.

Strongly held attitudes are also difficult to change because they shape how people interact with and interpret counter-attitudinal information. 
In some cases, people simply reject evidence that contradicts personal values or worldviews~\cite{lewandowskyMotivatedRejectionScience2016}.
But people need not reject evidence altogether for their attitudes 
to exert a pervasive, if more subtle, influence on how they consume information.
\revised{For instance, strong attitudes may lead to selective exposure, such that people avoid evidence that challenges their existing views, thereby limiting opportunities to be persuaded}~\cite{knobloch-westerwick_confirmation_2015}.
\revised{People also engage in motivated reasoning by selectively questioning the strength or validity of evidence that conflicts with their attitudes}~\cite{kraft_why_2015}.
Liao \textit{et al.}~\cite{liao2013beyond} examined how the personal relevance and motivation to learn about controversial issues (\emph{topic involvement}) impacted selective exposure and attitude change when exploring factual and opinion-based information.
Interestingly, in that study high topic involvement was associated with less selective exposure, such that people sought balanced sources of attitude-congruent and attitude-incongruent information \revised{for topics that were personally relevant}.
Yet participants rated attitude-congruent information more favorably and were less likely to change their attitudes compared to participants with low topic involvement~\cite{liao2013beyond}.

Beyond the strength of attitudes on a particular topic, researchers have identified individual factors related to the motivation and ability to engage with information that challenges existing views.
For instance, Albarracin \textit{et al.}~\cite{albarracinRoleDefensiveConfidence2004a} found that people who lack confidence in their ability to defend their attitudes (low \textit{defensive confidence}) prefer information that aligns with their existing views more than people with high defensive confidence.
Other work has shown that endorsement of scientific misconceptions and conspiracy theories are related to individuals' \textit{epistemic beliefs} about the nature of knowledge and how one justifies their own views~\cite{garrett_epistemic_2017,rudloffBeliefsNatureKnowledge2022}, including the importance of evidence, whether beliefs can be justified by intuition or gut feelings, and whether truth is ultimately determined by the sociopolitical context~\cite{garrett_epistemic_2017}.
Finally, recent work examining the susceptibility to misinformation indicates that the propensity to engage in analytic thinking is associated with more accurate judgements of the accuracy of fake news regardless of political attitudes, while people who rely on heuristic or intuitive thinking are more likely to believe misinformation that aligns with their political views~\cite{pennycookLazyNotBiased2019}.

In sum, while attitudes are informed by specific beliefs about empirical phenomena, they are also intertwined with personal values, identities, and ways of seeing the world. 
Whether data can change beliefs about controversial or personally relevant topics may depend on how those beliefs connect to broader attitudes, a question that we examine directly in our study by separately measuring the strength and certainty of participants' beliefs and attitudes related to the same topics.
We also measure individual characteristics previously linked to persuasion (defensive confidence, epistemic beliefs, and \revised{intuitive vs. analytic thinking}) to explore their relationship to topic-specific attitudes and beliefs.

\subsection{Impact of visualizations on belief updating and attitude change}

Visualization researchers have considered a broad range of strategies for effectively communicating data~\cite{franconeriScienceVisualData2021, viegas2006communication}, efforts which we see as typically intended to support \textit{belief change} about a quantity or relationship of interest. 
\revised{A growing number of studies have sought to directly measure this belief updating by assessing users' beliefs about a target claim or relationship before and after interacting with visualizations.}
For instance, Padilla \textit{et al.} examined how the CDC's COVID-19 forecast visualizations influence risk perception in two online experiments \cite{Padilla:2022:Nature}. They identified several factors that impact participants' perceived risks and their interpretation of the data, including the use of cumulative scales and the number of models displayed in the forecast visualization. \revised{These factors showed a varying degree of influence on whether people changed their beliefs about COVID-19 risks to oneself or others.}
In another recent example, Xiong \textit{et al.} used two crowdsourcing experiments to evaluate the effect of prior belief on interpretation of correlational relationships, showing that the statistical values people extract from the data can be biased by existing beliefs \cite{Xiong2022}.

In comparison to research at the intersection of data visualization and belief change, prior results are more mixed about whether data visualizations can induce  \textit{attitude change}. 
Heyer et al. \cite{heyer2020pushing} found that narrative visualizations prompted changes in attitudes and were more effective than text-based messages. 
However, eliciting prior beliefs (by having participants predict the data beforehand) did not significantly affect attitude change.
Liem et al. \cite{liem2020structure} found that visual narratives had small effects on attitudes about immigration, but that the effects differed according to demographic characteristics of the viewer.
Interestingly, Pandey \textit{et al.}~\cite{Pandey_TVCG_2014} found that persuasive messages with statistical evidence led to greater attitude change among people who did not have strong preexisting attitudes, and that among that group charts were more effective than tables. 
Notably, while these studies assessed changes in attitudes, they did not directly examine changes in beliefs about the message content and how belief updating related to global attitudes about the topic, a link that we aimed to examine in our study.

Another factor that may compound the resistance to belief and attitude change is \textbf{uncertainty representation} in data visualization. 
Researchers have questioned whether including uncertainty estimates in science communication undermines observers' trust in the evidence. 
van der Bles \textit{et al.} conducted five experiments in search of an answer to this question \cite{van_der_bles_effects_2020}. The findings revealed that communicating uncertainty numerically only produced a small negative effect on trust, which led to the recommendation for academics and science communicators to be more transparent about the limits of human knowledge. 
Similarly, Fischhoff \cite{Fischhoff2012CommunicatingUF} argued that communicating uncertainty is an author's moral imperative. 
However, Hullman \cite{Hullman8805422} found through surveys and interviews that data visualization authors worry that uncertainty obfuscates the signal a visualization expresses. 
While prior research on evaluating uncertainty encoding showed effects of different uncertainty representation on belief updating and decision outcomes~\cite{Fernandes2018, KaleKH21, Kale2019, karduni_bayesian_2021, WesslenKMD22} \revised{and produced a set of recommendations to reduce the mismatch between the conceptualization of uncertainty representation in visualization versus other fields \cite{hullman2018pursuit}}, more research is needed to understand the role of uncertainty representations in persuasive visualizations.

Our study design considers all three aforementioned factors: attitude change, belief change, and uncertainty representation to address a major research gap. 
More specifically, we evaluate changes in global topic-specific attitudes as well as beliefs about empirical relationships after seeing a series of visualizations with and without uncertainty representation related to a topic. 

\section{Current Study}

\revised{Consider this motivating} scenario: An individual with a strong anti-vaccine attitude observes data about vaccine effectiveness. 
\revised{Their anti-vaccine attitude might be linked to the belief that vaccines are harmful and can lead to increased fatality risk (i.e., that there is a positive correlation between vaccination rate and mortality rates).}
What should happen when they see a visualization depicting data that is incongruent with that prior belief, showing a strong negative correlation between vaccinations and deaths?
Assuming the data is credible, a rational observer would be expected to update their belief about the correlation to be closer to the observed data while accounting for the strength of their prior belief~\cite{karduni_bayesian_2021}. 
The viewer might be conservative when updating their beliefs if they have a strong prior, or if they engage in some form of motivated reasoning when interpreting the visualization.
Conservatism in belief updating might then be expected when people have especially strong priors (e.g., due to high familiarity or personal knowledge of a topic) or firmly established attitudes.

\revised{\subsection{Distinguishing beliefs and attitudes}
In the context of this study, we use the term \textit{belief} to refer to a person's conception of the relationship between two quantitative variables, in keeping with the broader definition of beliefs as an agreement with claims about the relationships between entities or variables. 
Belief updating is measured as the change in an individual's specific belief about that relationship after observing datasets depicted in scatterplot visualizations, which we treat as a proxy for an underlying learning process that unfolds as the person interacts with the data.
We also measure global \textit{attitudes} in order to capture an individual's general position on an issue (ranging from extremely against to extremely in favor).
We assess attitudes toward two topics (COVID-19 vaccination and labor union membership) and examine how they are related to specific beliefs about correlations between sets of variables.}

\begin{figure*}[t] 
\centering 
\includegraphics[width=\textwidth]{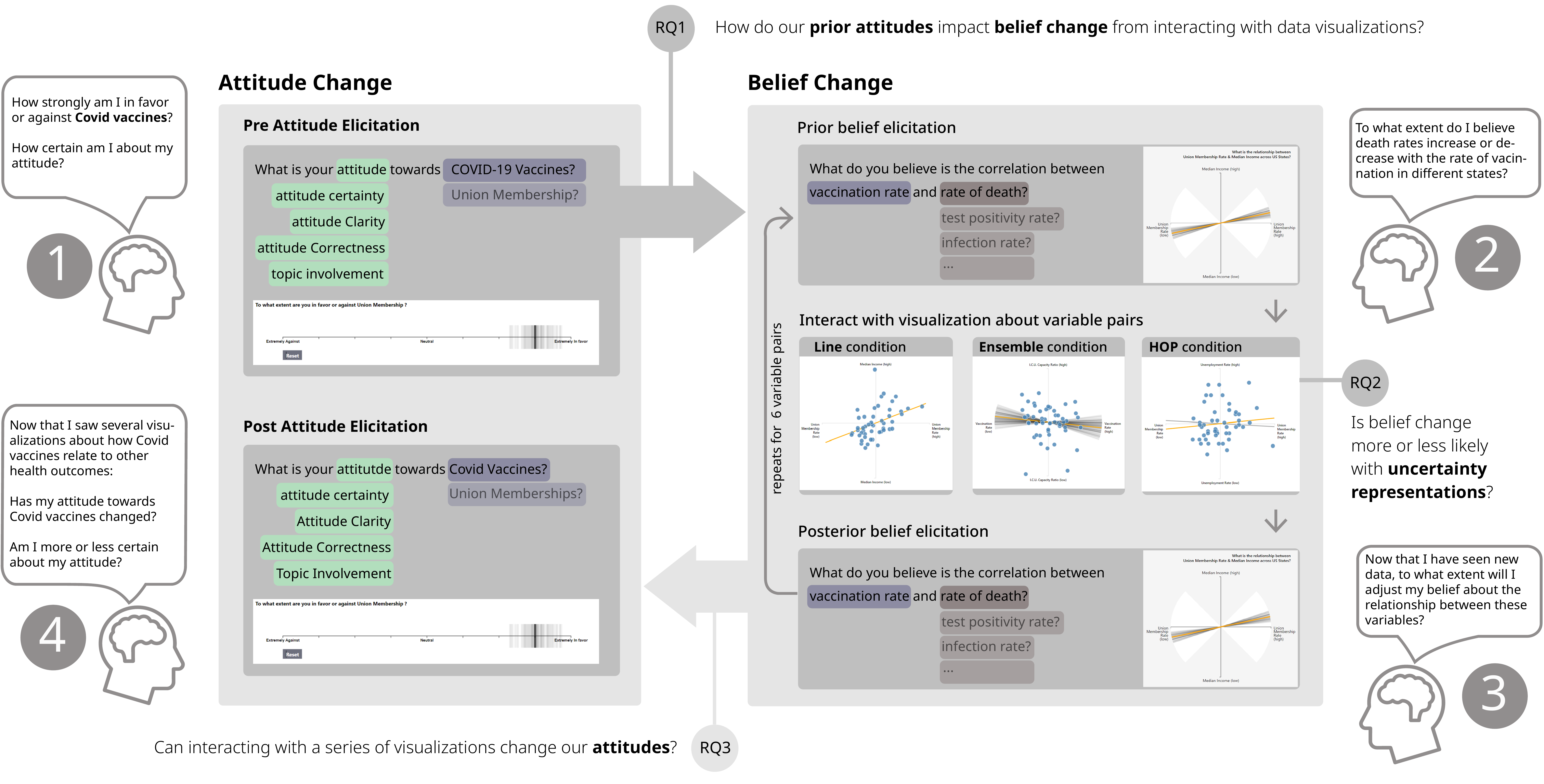}
  \caption{Design and flow of the present study examining how attitudes impact how people update their beliefs from data visualizations.}
  \Description{Figure shows the overall design of the present study examining how attitudes impact how people update their beliefs when interacting with visualizations. Participants report their overall attitude about a topic before completing a series of trials about related variable sets. For each trial, we elicit their beliefs about the correlation before and after they view the dataset. We also evaluate changes in attitudes that result from interacting with the datasets.}
\label{fig:design}
\end{figure*}

\subsection{\revised{Research questions}}

\revised{The first research question (RQ1)} is: How do global attitudes regarding a topic (e.g., positive or negative attitudes about vaccination) impact belief updating about specific empirical relationships (e.g., the association between vaccination and mortality rates)? In our study we chose scatterplots as the visual encoding of bivariate relationships. 
Scatterplots are one of the simplest and most common forms of data visualizations in which data points are shown on a two-dimensional plot. 
A common use of this representation is to convey the correlation between two variables; that is, how differences in one variable are related to the direction and magnitude of differences in the other variable.

The second research question \revised{(RQ2) concerns the role of uncertainty representations in belief and attitude change.} 
We designed three conditions with and without uncertainty representation of a correlation that accompanied each scatterplot.
The \textit{Line} condition served as a baseline in which the best-fit correlation line was superimposed on the data points.
In the hypothetical outcome plot (\textit{HOP}) condition~\cite{Kale2019}, the baseline visualization was augmented with animated draws from the 95\% confidence interval for the correlation based on the dataset.
In the \textit{Ensemble} condition, those draws were displayed in a static representation.
Thus, the HOP and Ensemble conditions conveyed the same uncertainty but using dynamic or static representations, respectively.
We predicted that visualizations with uncertainty representations (HOP and Ensemble conditions) would lead to less belief updating when the data was incongruent with prior beliefs and attitudes. 

The third research question focuses on attitude changes after interacting with a series of visualizations for a specific topic \revised{(RQ3)}. In addition to examining how existing attitudes affect belief updating, we also evaluated attitude change after participants interacted with a series of six visualizations related to each topic.
We were interested in whether attitudes would shift as a result of interacting with data that may conflict with prior beliefs.

\section{Experiment}
To investigate our research questions we conducted an online experiment with participants from Prolific with a task that follows the design flow shown in Figure \ref{fig:design}.
\revised{We preregistered our study design, exclusion criteria, main hypotheses related to our research questions, and overall analysis approach} at the following link: \url{https://aspredicted.org/5pp28.pdf}. 
\revised{Deviations from the preregistration are noted in the relevant sections below.}

\subsection{Topics and Datasets}
\begin{figure*}[bt] 
\centering 
\includegraphics[width=6in]{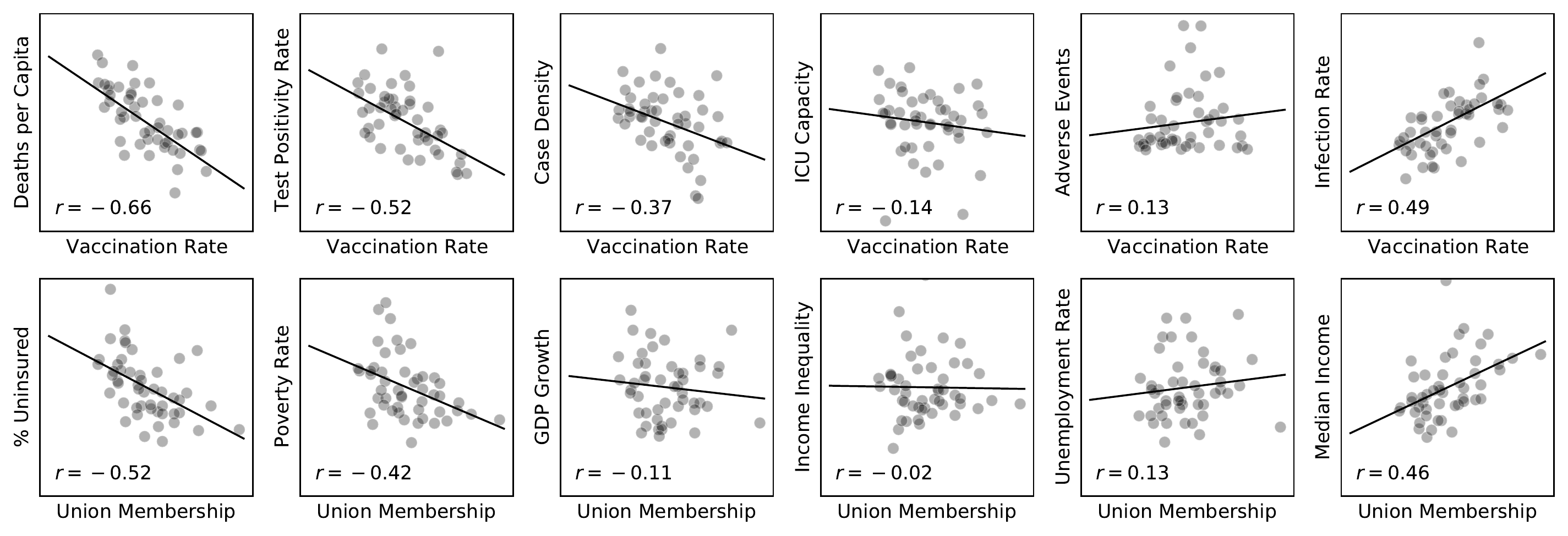}
\caption{Datasets for COVID-19 vaccination topic (top row) and labor union topic (bottom row).} 
\label{fig:datasets}
\Description{Figure shows the 12 datasets presented to participants in the experiment. Each dataset is represented as a scatterplot with a correlation line depicting the relationship between a focal variable (e.g., COVID-19 vaccination) and another outcome.}
\end{figure*}

\revised{We selected two topics, COVID-19 vaccination and labor union membership, for the current study because we expected participants to differ in the strength of preexisting attitudes on these two topics. 
This choice was informed by a previous study by Karduni \textit{et al.} in which bivariate datasets on these topics were included but with artificial data ~\cite{karduni_bayesian_2021}. 
In that study the authors found a weak influence of prior beliefs on belief updating, potentially because the experiments relied on fabricated datasets of an unknown provenance. 
In the present study, we use real data obtained from public sources and show information about those sources to participants to enhance the realism and credibility of the data visualizations. }
Figure \ref{fig:datasets} shows the 6 datasets for each topic (Covid-19 vaccination and labor union membership) that the participants interacted with during the experiment.

\textbf{COVID-19 vaccination.} COVID-19-related data were retrieved using the COVID-19 Act Now data API \cite{covidActNow}.
Infection rate, and case density data are originally from the New York Times \cite{newYorkTimesApi}, while ICU capacity ratio \cite{HHSICUcapacity}, and test positivity ratio \cite{HHSTestpositivityratio} were from the Department of Health and Human Services. Adverse vaccination events data \cite{CDCAdverseVaccinationEvents} were from Centers for Disease Control and Prevention (CDC) and vaccination rate were from the CDC \cite{CDCVaccinationRate} and North Carolina Department of Health and Human Services COVID-19 Response \cite{NCHHSVaccinationRate}.
All variables were averaged over the 2021 calendar year, the period during which vaccines became widely available in the United States.

\textbf{Labor union membership.} Data for labor union membership in 2019 was retrieved from ~\cite{hirsch2003union}. 
State-level data for GPD growth, median income, unemployment rate, percent of population without insurance, poverty rate, and income inequality were obtained from the 2019 US Census Current Population Survey and American Community Survey \cite{USCensusPSAS}.

\textbf{Data preparation.} All variables were normalized by z-scoring.
Scatterplots of the resulting datasets with  correlation lines are shown in Figure~\ref{fig:datasets}.
In selecting variable sets, we sought to include a range of correlations with the focal variable for each topic (COVID-19 vaccination rate or the proportion of union membership by state) and to include some variable sets that were consistent with positive or negative attitudes.
For example, for the COVID-19 data, the negative relationship between vaccination rate and deaths per capita is consistent with a pro-vaccination attitude, while the positive relationship between vaccination rate and infection rate might be interpreted as supporting an anti-vaccination attitude. 
Similarly, for the union data, the negative relationship between union membership rate and the proportion of uninsured is consistent with a pro-union attitude, while the weakly positive relationship between union membership and unemployment rate might be seen as consistent with an anti-union attitude.

\subsection{Participants}

We recruited 412 participants from Prolific. Participants earned
\$4~upon completion of the task, which took an average of 26 minutes (SD
= 12.5). Per our preregistration, we excluded \textcolor{black}{81}
participants who did not complete the full study. We planned to exclude
any participants who gave non-sensical or inappropriate responses to the
open-ended questions for the CRT, but no participants met this
criterion.
\textcolor{black}{Based on initial pilot testing we also preregistered exclusion criteria based on task duration, such that we planned to exclude participants who completed the study is less than 15 minutes or more than 50 minutes. However, this led to the exclusion of an unexpectedly large number of additional participants ($n$ = 67). Further inspection of these participants' data did not reveal other indications of low-effort or inappropriate responses, suggesting that our initial criteria for task duration were too strict. We therefore decided to remove this exclusion criterion for analyses reported below. However, we also performed the same analyses with the original exclusion criteria in place and found there to be no substantive differences that would impact our conclusions.}

After \textcolor{black}{excluding incomplete responses}, there were
\textcolor{black}{$N$ = 331} participants included in the analyses
\textcolor{black}{(Line: 108; Ensemble: 113; HOPs: 110).} The average age
was \textcolor{black}{40.7} years (\(SD\) = \textcolor{black}{13.5}, range:
\textcolor{black}{19--73}). \textcolor{black}{165} identified as female,
\textcolor{black}{157} identified as male, \textcolor{black}{7} as another
gender, \textcolor{black}{2} chose not to respond. \textcolor{black}{260}
were White, \textcolor{black}{26} were Black/African-American,
\textcolor{black}{28} were Asian, \textcolor{black}{2} were American Indian
or Alaska Native, \textcolor{black}{12} were another race/ethnicity,
\textcolor{black}{3} gave no response. Most participants had obtained a
college degree and the sample covered a range of political affiliations,
with \textcolor{black}{53}\% liberal, \textcolor{black}{19}\% moderate, and
\textcolor{black}{27}\% conservative (see full distribution of responses
in Supplementary Figure \ref{fig:demographics}).

\subsection{Design and Procedure}

We employed a mixed design with a between-subjects manipulation of the visualization type (see \ref{VisConditions}) and a within-subjects manipulation of the topics presented to participants (COVID-19 vaccination and union membership). 
The main task had two rounds corresponding to the two topics (Figure~\ref{fig:design}).
In each round, participants first responded to questions assessing their global attitude, attitude certainty (clarity and correctness) and involvement.
After reporting their global attitude, participants completed six visualization trials for each topic, with each trial comprised of a 1) prior belief elicitation, 2) interaction with a data visualization, and 3) posterior belief elicitation.
After completing all six trials for the topic, participants again responded to questions about their global attitude, attitude certainty, and topic involvement.
The order of topics and datasets within each topic were randomized for each participant.

\revised{ \textit{Visual elicitation methods.} The experiment leveraged a graphical elicitation approach developed in two previous studies to measure both the strength and uncertainty of people's beliefs and attitudes ~\cite{karduni_bayesian_2021, karduni2023images}. For belief elicitation, we used the Line+Cone interface developed by Karduni \textit{et al.} to visually elicit beliefs about correlations  
when viewing bivariate data with different sample correlations~\cite{karduni_bayesian_2021}. 
\revised{In that study the authors compared the Line+Cone method to a more labor-intensive MCMC-P method for estimating beliefs and found that Line+Cone achieved comparable results. The authors also included comprehension tests showing that people understood the visual representation of correlation.}
Both prior and posterior beliefs were elicited before and after seeing a bivariate correlation visualization, as we are interested in how participants update their beliefs based on their prior knowledge and the observed data. 
For attitude elicitation, we used the visual attitude scale proposed by Karduni et al. \cite{karduni2023images} that allows participants to specify the value and uncertainty range of their attitudes. 
Both belief and attitude elicitation methods are shown in Figure \ref{fig:design}. }

\subsubsection{Eliciting global topic attitudes and involvement}
\label{Attitude_measure}

\textbf{Attitude valence.} In each round, participants reported their global attitude in relation to the topic (i.e., the extent to which they support or oppose policies that cause people to get vaccinated or to become members of unions).
Responses were made on a continuous scale with the endpoints labeled ``Extremely against'' and ``Extremely in favor.'' 
Participants recorded the overall valence of their attitude (\(a\)) by clicking on the scale.
They then adjusted an interval (\([a_{upper}, a_{lower}]\)) to indicate their uncertainty about their attitude towards the topic, with larger intervals indicating a greater degree of uncertainty.
Attitudes were elicited both before and after seeing all data visualizations for the topic, resulting in 6 measurements for each topic (\(a^{prior}\), \(a^{prior}_
{upper}\), and \(a^{prior}_{lower}\),  \(a^{post}\), \(a^{post}_{upper}\), and \(a^{post}_{lower})\). 


\textbf{Attitude clarity and correctness.} \revised{Since the visual attitude elicitation method was only recently developed, we also included} questions from Petrocelli \textit{et al.}~\cite{Petrocelli2007} to independently assess two dimensions of attitude certainty: clarity and correctness.
\textit{Attitude correctness} refers to feeling that an attitude is correct or justified.
This scale includes three questions about whether a person believes that their attitude is the ``correct attitude,'' the ``right way to think and feel about the issues'' and that ``other people should have the same attitude.'' 
\textit{Attitude clarity} concerns whether one feels confident in their ability to report their own attitude on a topic.
This scale includes four questions about whether the stated attitude reflects a person's ``true thoughts and feelings,'' is ``clear in your mind,`` and ``is really the attitude you have.''

\textbf{Topic involvement.} \revised{Since prior research has found that topic involvement impacts the evaluation of evidence that is inconsistent with an preexisting attitude}, we asked the participants to respond to four questions from Liao and Fu~\cite{liao2013beyond} to assess the self-relevance and motivation to learn about each topic, with 5 point scale from ``not at all'' to ``extremely.'' 
The items were: ``To what extent is this topic related to your core values?'', ``To what extent is it important for you to defend your point of view on this topic?'', ``How interested are you in learning about this topic?'', and ``To what extent are you motivated to know the truth about this topic?''

\subsubsection{Eliciting beliefs about correlations}

Beliefs about each variable pair were elicited before and after participants interacted with a data visualization.
The Line + Cone interface~\cite{karduni_bayesian_2021} was designed to quickly elicit beliefs about a bivariate correlation using a two-step interaction (Figure~\ref{fig:design}, right).
First, the participant adjusts the angle of a line to indicate what they believe is the most likely relationship (\(r\)) between the two variables, ranging from a perfect negative correlation at –45 degrees ($r = -1$) to a perfect positive correlation at +45 degrees ($r = +1$).
After making a selection, the participant adjusts the size of a cone of uncertainty.
The cone is represented by an ensemble of gray lines that are drawn from a Normal distribution centered on the most likely correlation and truncated at -1 and +1.
Participants modify the spread of the lines to cover an interval ($CI_{elicited}$: \([r_{upper}, r_{lower}]\)) of ``plausible alternatives'' for the relationship, with larger intervals indicating more uncertainty about the true relationship.
Participants were self-paced and could reset the interface in order to change their response.

\revised{Before the first trial participants completed a short training on the belief elicitation interface. 
These instructions defined how a relationship between two variables is represented by a line and how the slope of the line represents different relationships (positive, neutral, or negative). 
An example showing the relationship between Height and Weight was also provided. 
In the next step, participants were provided written instructions on how to respond with the belief elicitation tool and were tested on their understanding of the relationship between two variables.}

Beliefs were elicited both before and after seeing the dataset, resulting in 6 measurements for each variable set (prior: \(r^{prior}\), $CI^{prior}$; posterior: \(r^{post}\), $CI^{post}$). 
For the prior belief elicitation participants were asked to ``consider what you think is the true relationship between the following variables.''
For the posterior belief elicitation participants were instructed as follows: ``Now that you have seen a dataset involving these variables, consider again what you think is the true relationship between the variables. Note that we are interested in your genuine beliefs about this relationship, whether or not they were affected by the data shown on the last page.''
\revised{These instructions were meant to discourage participants from responding to the posterior elicitation by simply recreating the correlation shown in the preceding scatterplot.}

Based on the elicited beliefs before and after each dataset we calculated difference scores for the change in beliefs about the most likely relationship (\(r^{post} - r^{prior}\)) and change in uncertainty ($CI^{post} - CI^{prior}$).
\revised{We also evaluated the difference between participants' beliefs about the correlation at each stage and the actual correlation shown in the scatterplot:}
We used the true sample correlation \(r_{data}\) for a given dataset (shown in Figure~\ref{fig:datasets}) to calculate the prior belief distance (\(|r_{data} - r^{prior}|\)) and posterior belief distance (\(|r_{data} - r^{post}|\)).

\subsubsection{Data visualization conditions}
\label{VisConditions}
\revised{Since RQ2 aims to investigate the role of uncertainty visualization in belief and attitude change, our experiment included three visualization conditions with and without uncertainty representation.} In each trial, after completing the prior belief elicitation, participants viewed the dataset for the current variable pair. Participants were randomly assigned to one of three visualization conditions (see examples in Figure~\ref{fig:design}):

\begin{itemize}
    \item \textbf{Line:} The Line condition served as the baseline with a scatterplot and a superimposed correlation line. This condition contains no uncertainty representation.
    \item \textbf{HOP:} Hypothetical outcome plots (HOPs) were used to present animated draws from the 95\% confidence interval for the population correlation based on the dataset. 
    \item \textbf{Ensemble:} The ensemble display was a static representation of the same random draws from the 95\% confidence interval from the population correlation as in the HOP condition.
\end{itemize}

The choice of the three visualization conditions were informed by previous studies on belief change in the context of uncertainty visualization \cite{karduni_bayesian_2021,WesslenKMD22}. The correlation line was present in all conditions, and participants were instructed that ``The orange line indicates the estimated relationship between the two variables based on this dataset.'' Participants in the HOP and Ensemble conditions were also told that ``the gray lines indicate the uncertainty in the relationship between the variables, with each line representing a plausible alternative relationship.''

Participants could interact with the visualization by hovering over a point, which revealed a label indicating the US state and precise values of the variables for that datapoint.
Alongside the dataset we provided textual descriptions of the meaning of the variables, information about data sources, \revised{and generic instructions on how to interpret the plots.}
Participants were self-paced when viewing each dataset and were not required to make any responses other than pressing a button to continue to the posterior elicitation.

\subsection{Questionnaires}
\revised{We included several additional questionnaires that evaluated individual factors related to the motivation and ability to engage with information that challenges existing views \cite{albarracinRoleDefensiveConfidence2004a, garrett_epistemic_2017,rudloffBeliefsNatureKnowledge2022} as reviewed in section 2.1.
The inclusion of these measures was primarily exploratory (with the exception that we also planned to use the Cognitive Reflection Test as an attention check).
We provide an analysis of the relationships between topic-specific attitudes and these individual measures in Supplemental Materials Section~\ref{supp:covariates}, but since we did not have \textit{a priori} hypotheses about these measures related to our study design we do not analyze them further.
}

\textbf{Epistemic beliefs.} We used the questions from~\cite{garrett_epistemic_2017} to measure epistemic beliefs along three dimensions: \textit{Faith in intuition for facts} (e.g., ``I trust my gut to tell me what’s true and what’s not''), \textit{Need for evidence} (e.g., ``I need to be able to justify my beliefs with evidence''), and \textit{Truth is political} (e.g., ``Facts are dictated by those in power'').
There were four items for each dimension, with responses on a 5-point scale from ``strongly disagree'' to ``strongly agree.''

\textbf{Defensive confidence.} The defensive confidence scale~\cite{albarracinRoleDefensiveConfidence2004a} is comprised of twelve items assessing the extent to which people feel they can defend their attitudes (e.g., ``During discussions of issues I care about, I can successfully defend my ideas.''). Participants responded on a 5-point scale from ``not at all characteristic of me'' to ``extremely characteristic of me.''

\textbf{Cognitive Reflection Test (CRT).} We used the 4-item Cognitive Reflection Test from \cite{thomson_investigating_2016} to measure the propensity to engage in intuitive vs. analytic thinking.
The CRT includes four free-response items, each of which has an ``intuitive'' response that differs from the correct, ``analytical'' response (e.g., ``If you’re running a race and you pass the person in second place, what place are you in?''; intuitive response: first; correct answer: second).

The order of the Epistemic beliefs and Defensive confidence questionnaires were randomized for each participant, with one occurring before the main task and the other at the end of the session.
The four CRT questions were interspersed with other questionnaires, with two questions in the beginning and two at the end of the session.

\section{Results}

\begin{figure*}[bt]
\centering
\includegraphics[width=.9\textwidth]{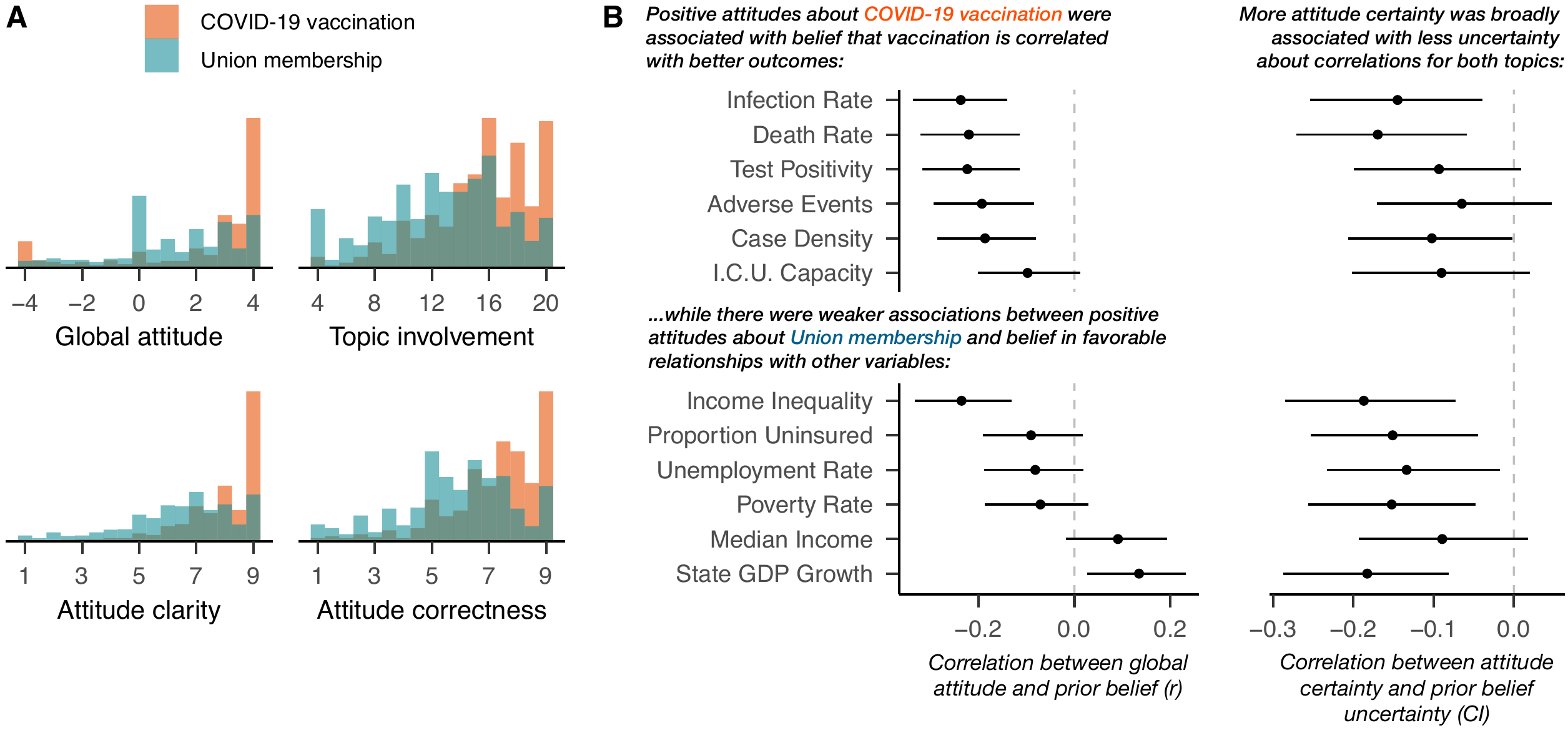}
\caption{\textbf{A:} \revised{Histograms of responses} for global attitude, topic involvement, attitude certainty (clarity and correctness) for each topic, assessed before interacting with the data visualizations. \textbf{B:} Estimated correlations \revised{(posterior means and 95\% credible intervals) between topic-specific attitude valence and prior belief (left) and between topic-specific attitude certainty and prior belief uncertainty (right) for each variable pair.}}
\label{fig:attitudes}
\Description{See description in text.}
\end{figure*}

\revised{All analyses were conducted in R version 4.1.2~\cite{RManual}. 
Data and analysis code are available at \url{https://osf.io/wvqky/}.
We performed Bayesian statistical analyses and report posterior means and 95\% credible intervals for all statistics, as well as Bayes Factors (BF) for hypothesis tests unless otherwise noted.
For tests reported below, the Bayes Factor represents the strength of evidence in favor of a relationship compared to null hypothesis, with $BF > 1$ indicating evidence in favor of an effect and $BF < 1$ indicating evidence in favor of the null \cite{heck2022review}.
Correlations were estimated using the \textit{correlation} library \cite{Rcorrelation2020} with a null hypothesis of $r = 0$ and default priors.  
Two-group (within-subjects) comparisons were conducted using \textit{brms} \cite{burkner2017brms} to perform robust estimation with the difference between measurements modeled with a t-distribution as described in \cite{kruschkeBayesianEstimationSupersedes2013}, with a null hypothesis corresponding a mean difference of 0. 
}

\subsection{Prior attitudes and beliefs}

We first conducted non-preregistered exploratory analyses to examine the relationships between topic-specific attitudes and prior beliefs about the variable sets.

\textbf{Global attitude valence and certainty.} Distributions of
dependent measures related to attitudes about each topic are shown in
Figure \ref{fig:attitudes}. Global attitudes toward COVID-19 vaccination
were mostly favorable (\(M\) = \textcolor{black}{2.24}, \(SD\) =
\textcolor{black}{2.52}), with a smaller number of participants who were
strongly opposed. Attitudes toward union membership were also positive
overall (\(M\) = \textcolor{black}{1.26}, \(SD\) = \textcolor{black}{2.1})
but more variable, with many participants expressing neutral or negative
attitudes. Attitudes toward union membership were less positive compared
to COVID-19 vaccination
\textcolor{black}{(mean difference $D$ = 1.08 [0.82, 1.34], $BF$ = \ensuremath{2.96\times 10^{11}})}.

Participants' confidence about their attitudes also differed between the
two topics, with lower ratings of attitude clarity
\textcolor{black}{(mean difference $D$ = 1.36 [1.15, 1.57], $BF$ = \ensuremath{1.89\times 10^{36}})}
and attitude correctness
\textcolor{black}{(mean difference $D$ = 1.4 [1.17, 1.61], $BF$ = \ensuremath{3.24\times 10^{29}})}
for union membership compared to COVID-19 vaccination. Participants also
reported lower topic involvement for union membership compared to
COVID-19 vaccination
\textcolor{black}{(mean difference $D$ = 2.79 [2.36, 3.22], $BF$ = \ensuremath{1.19\times 10^{29}})}.
Stronger attitudes about either topic were also held with more
confidence, as attitude strength (the absolute value of prior attitude)
was positively correlated with attitude clarity
\textcolor{black}{(COVID-19: Spearman $r$ = 0.57 [0.51, 0.65], $BF$ = \ensuremath{1.25\times 10^{28}}; Union: $r$ = 0.59 [0.53, 0.66], $BF$ = \ensuremath{1.4\times 10^{30}}) and attitude correctness (COVID-19: $r$ = 0.53 [0.46, 0.61], $BF$ = \ensuremath{2.13\times 10^{23}}; Union: $r$ = 0.6 [0.53, 0.67], $BF$ = \ensuremath{3.36\times 10^{31}}), and topic involvement (COVID-19: $r$ = 0.31 [0.21, 0.4], $BF$ = \ensuremath{1.52\times 10^{6}}; Union: $r$ = 0.5 [0.41, 0.57], $BF$ = \ensuremath{5.18\times 10^{19}}).}
These correlations are consistent with existing literature on attitudes,
in that people tend to be more certain about attitudes that are more
extreme or personally important
\cite{Petrocelli2007,tormalaAttitudeCertaintyAntecedents2018}.

We also directly measured attitude uncertainty using the graphical
elicitation method described in section \ref{Attitude_measure}. Based on
prior work we expected this direct measure of attitude uncertainty to be
negatively correlated with attitude clarity and correctness, but we
instead found small positive correlations between these measures. Upon
closer examination of responses we found that a group of participants
had large attitude uncertainty ranges (e.g., ranges spanning the whole
spectrum of possible attitudes) but also reported high clarity and
correctness scores, suggesting these participants may have misunderstood
the graphical elicitation technique. Therefore, in subsequent analyses
we use the mean of attitude clarity and correctness (which were strongly
positively correlated: Spearman \(r\) =
\textcolor{black}{0.74 [0.7, 0.77], $BF$ = \ensuremath{10^{113}}}) as a
combined index of attitude certainty.

\begin{figure*}[t]
\centering
\includegraphics[width=.9\textwidth]{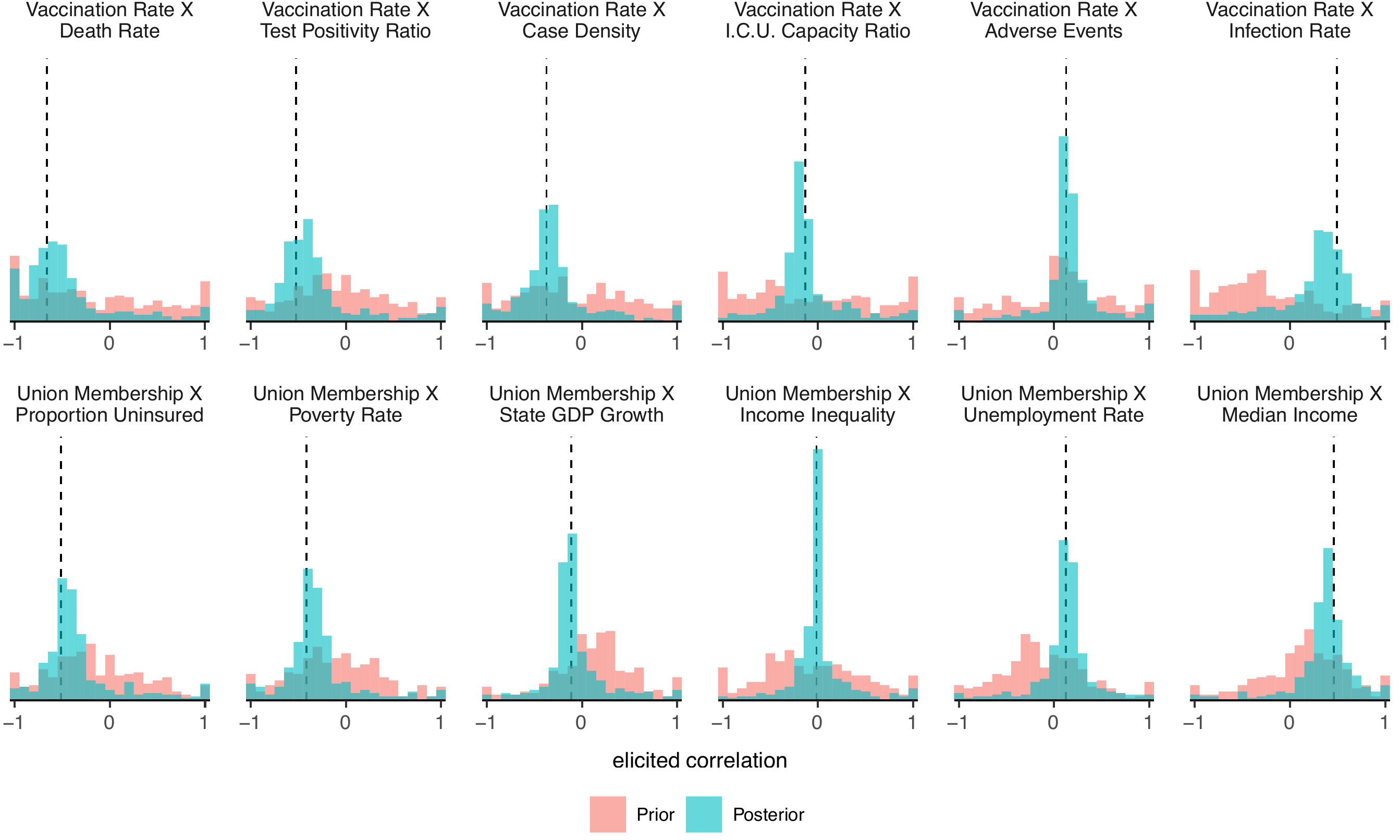}
\caption{\revised{Histograms} of prior and posterior beliefs for each variable pair. Vertical dashed lines show the sample correlation for each dataset.}
\label{fig:beliefs_prepost}
\Description{The figure shows the distributions of responses to the prior and posterior belief elicitation for each variable set. In general, posterior beliefs converged toward the true sample correlation.}
\end{figure*}

\textbf{Global topic attitudes and prior beliefs about correlations.} We next examined the relationship between global topic-specific attitudes and prior beliefs about specific variable pairs.
\revised{As noted above, prior research on attitudes indicates that they entail associations with specific beliefs about observable phenomena. In the context of our study, we would expect that participants with positive attitudes about an action (e.g., COVID-19 vaccination) would tend to believe that action is associated with favorable outcomes (e.g., lower fatality rates when the vaccination rate is high). 
To evaluate this connection between prior attitudes and beliefs, we estimated the correlations between topic-specific attitude (\(a^{prior}\)) and the prior belief about the most likely relationship (\(r^{prior}\)). Figure \ref{fig:attitudes}B (left) shows the posterior mean and 95\% credible intervals for the correlations for each variable pair.}
For COVID-19 variables, participants who had positive attitudes toward vaccination also predicted more negative relationships between vaccination rate and other outcomes\revised{---in other words, that increased vaccination would be related to reduced risk of infection, fatality, etc.}
\revised{A similar pattern emerged for union membership, but for that topic the relationships between attitudes and prior beliefs varied more across variable sets. People with pro-union attitudes tended to think union membership was linked to less income inequality and higher GDP growth, while beliefs about other outcomes were less closely tied to global attitudes about union membership. This suggests that, in contrast to COVID-19 vaccination, attitudes about union membership were not associated with strong prior beliefs about its relationship to some of the socioeconomic variables included in the study.}

\revised{We conducted a similar analysis of the relationship between topic-specific attitude certainty and participants' uncertainty about the correlations between a pair of variables, as measured by the size of the elicited CI. In general, higher attitude certainty was negatively related to belief uncertainty, such that people who were more confident about their own attitude tended to make more precise prior estimates of the correlation between variables (Figure \ref{fig:attitudes}B, right).}


In sum, participants' views about the two topics differed in the strength and confidence of their attitudes, and these global attitudes were related to their prior beliefs about specific empirical relationships. 
Attitudes about COVID-19 vaccination were more strongly polarized, and pro-vaccination attitudes were associated with the belief that higher vaccination rates would correlate with favorable health outcomes.
Within the same group of participants, attitudes toward union membership were more ambivalent and less predictive of specific beliefs about how union membership relates to certain socioeconomic outcomes, suggesting participants had less well-formed beliefs about those variable pairs compared to COVID-19 vaccination.
\revised{In the next section we examine whether these differences in prior views toward each topic impacted how people updated their beliefs about specific empirical relationships when viewing scatterplot visualizations.}

\subsection{Change in beliefs about the most likely correlation \revised{(RQ1\&2)}}

Beliefs about the most likely correlation between each pair of variables were elicited both before and after interacting with each data visualization.
The distributions of prior ($r^{prior})$ and posterior beliefs ($r^{post}$) about the most likely correlation are shown for each variable pair in Figure~\ref{fig:beliefs_prepost}. 
In the aggregate, there were marked shifts toward the true sample correlations of each dataset (vertical dashed lines). 
\revised{In this section we examine whether these changes in beliefs were related to individuals' topic-specific attitudes and/or the presence of uncertainty representations in the scatterplot visualizations.}

\begin{figure*}[t] 
\centering 
\includegraphics[width=6.8in]{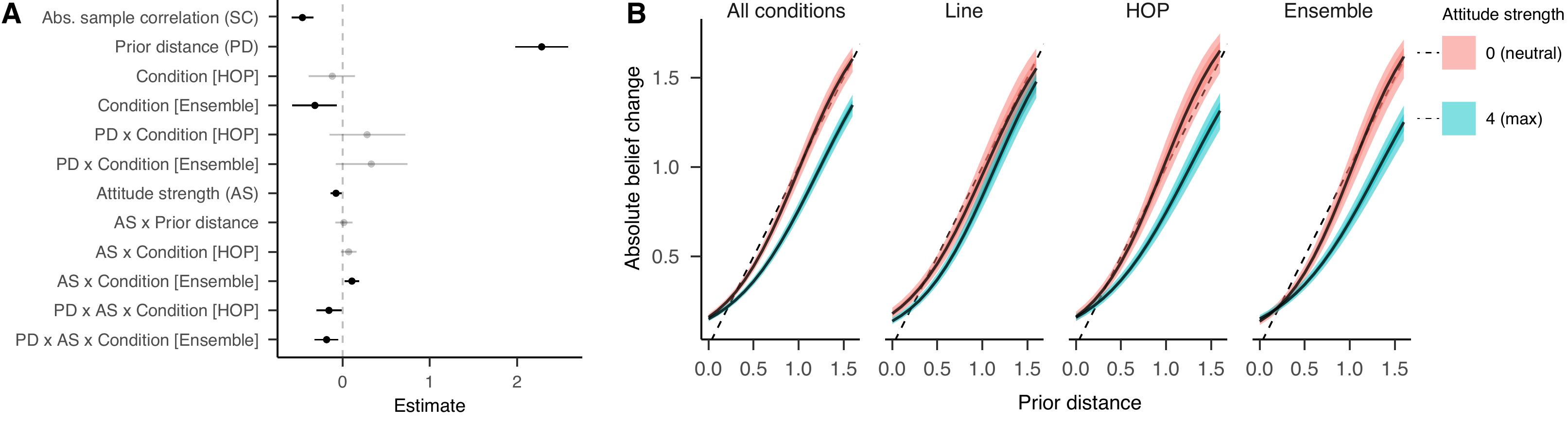}
\caption{\revised{\textbf{A:} Estimated fixed effects (posterior mean and 95\% credible intervals) for model of absolute belief change. 95\% credible intervals that include zero are in gray. \textbf{B:} Expected predictive distributions for absolute belief change by visualization condition and global attitude strength. The figure depicts a contrast between the model predictions when attitude strength is 0 (neutral attitude toward a topic) compared to when attitude strength is 4 (the maximum rating).}} 
\label{fig:belief_models}
\Description{The figure shows the results of the absolute belief change model (see full description in text).}
\end{figure*}

\subsubsection{Absolute belief change}
\label{Belief_change}

In our first preregistered analysis of belief change we used Bayesian multilevel models to examine whether the amount of belief change was affected by \revised{global topic attitudes (RQ1) or uncertainty representations (RQ2)}.
Using the \textit{brms} \textit{R} package~\cite{burkner2017brms}, we performed zero-one-inflated-Beta regression to model absolute belief changes about the most likely correlation (transformed to range from 0--1: \(|r^{post} - \revised{r^{prior}}|/2\)) with random intercepts for participants and fixed effects for visualization condition (Line/HOP/Ensemble).
\revised{All models also included fixed effects for (1) the prior distance (the absolute difference between the sample correlation and the participant's prior belief), since we did not expect to observe any belief change when the prior belief already matched the sample correlation (i.e., when prior distance = 0); and (2) the absolute value of the sample correlation, as we expected belief changes to be larger when the sample correlation was more extreme.}
We compared this baseline set of predictors to models that \revised{included fixed effects for} global attitude strength (the absolute value of global attitude for each topic) and its interaction with prior distance and visualization condition.\footnote{\revised{In our preregistration we specified a simpler baseline model which only included a fixed effect for visualization condition, but we opted to also include prior distance and absolute sample correlations in the baseline model based on the results of~\cite{karduni_bayesian_2021} which showed strong effects of both factors on belief change.}}
Models were evaluated using Pareto smoothed importance sampling to approximate
leave-one-out cross validation (PSIS-LOO), which estimates the expected log predictive density, or the model's expected ability to predict new data~\cite{vehtari_practical_2017-1}. 
Full details of the model definitions and results of the model comparison are in Supplementary Materials Section~\ref{ABC_model_supplement}.


The best-performing model (ABC-10) included \revised{credible effects of visualization condition, prior distance, and attitude strength, and interactions between them} (see \revised{Figure~\ref{fig:belief_models}A for posterior means and 95\% CIs for the main predictors of interest} and the full table of estimated coefficients in Supplementary Table S5).
\revised{The model is summarized in the expected posterior predictions shown in Figure~\ref{fig:belief_models}B, which shows the predicted absolute belief change as a function of prior distance for two attitude strengths: a neutral attitude (attitude strength = 0) and a strong attitude (attitude strength = 4).
As expected, there was a large positive effect of prior distance on absolute belief change: When prior distance was low people were unlikely to change beliefs that already matched the sample correlation conveyed by the visualization, while absolute belief change increased for larger prior distances where the sample correlation differed from the prior belief.}

\revised{If participants shifted their posterior belief to match the observed sample correlation in a given dataset, then the amount of belief change would follow the diagonal dashed lines in Figure~\ref{fig:belief_models}B.}
However, the best-performing model also indicated credible interactions between prior distance and attitude strength.
Neutral attitudes about a topic (attitude strength of 0) were associated with changes in beliefs in proportion to the prior distance (illustrated by the red ribbon following the dashed line), \revised{such that people without strong existing attitudes for a topic were willing to make large adjustments when their prior belief conflicted with the data}.
In contrast, strong global attitudes (attitude strength of 4) were associated with smaller changes in beliefs about the most likely relationship \revised{(shown by the blue ribbon that falls below the dashed line in Figure~\ref{fig:belief_models}B)}.
\revised{This provides strong evidence that, despite viewing the same datasets, participants with strong preexisting attitudes about a topic made smaller adjustments to their beliefs about a given correlation compared to participants with a neutral attitude.}

Interestingly, the best-performing model also included credible interactions between between prior distance, attitude strength, and visualization condition, such that \revised{the effect of global attitude was more pronounced in visualization conditions with uncertainty representations (HOP and Ensemble) compared to visualizations with only the scatterplot and correlation line (Line condition) (see Figure~\ref{fig:belief_models}B). This result is consistent with our hypothesis for RQ2 that uncertainty representations would lead to less belief updating when prior views conflict with the data.}
However, it's important to note that alternative models achieved comparable performance according to the LOO criterion (Supplementary Table S4).
\revised{Under these alternative models there were consistent effects of attitude strength on absolute belief change, but some did not include credible interactions between visualization condition and attitude strength. 
Thus, we suggest that the difference between visualization conditions, such that uncertainty representations produce smaller changes in beliefs when people have strong attitudes, warrants further investigation to evaluate its robustness}.

\begin{figure*}[t] 
\centering 
\includegraphics[width=5.5in]{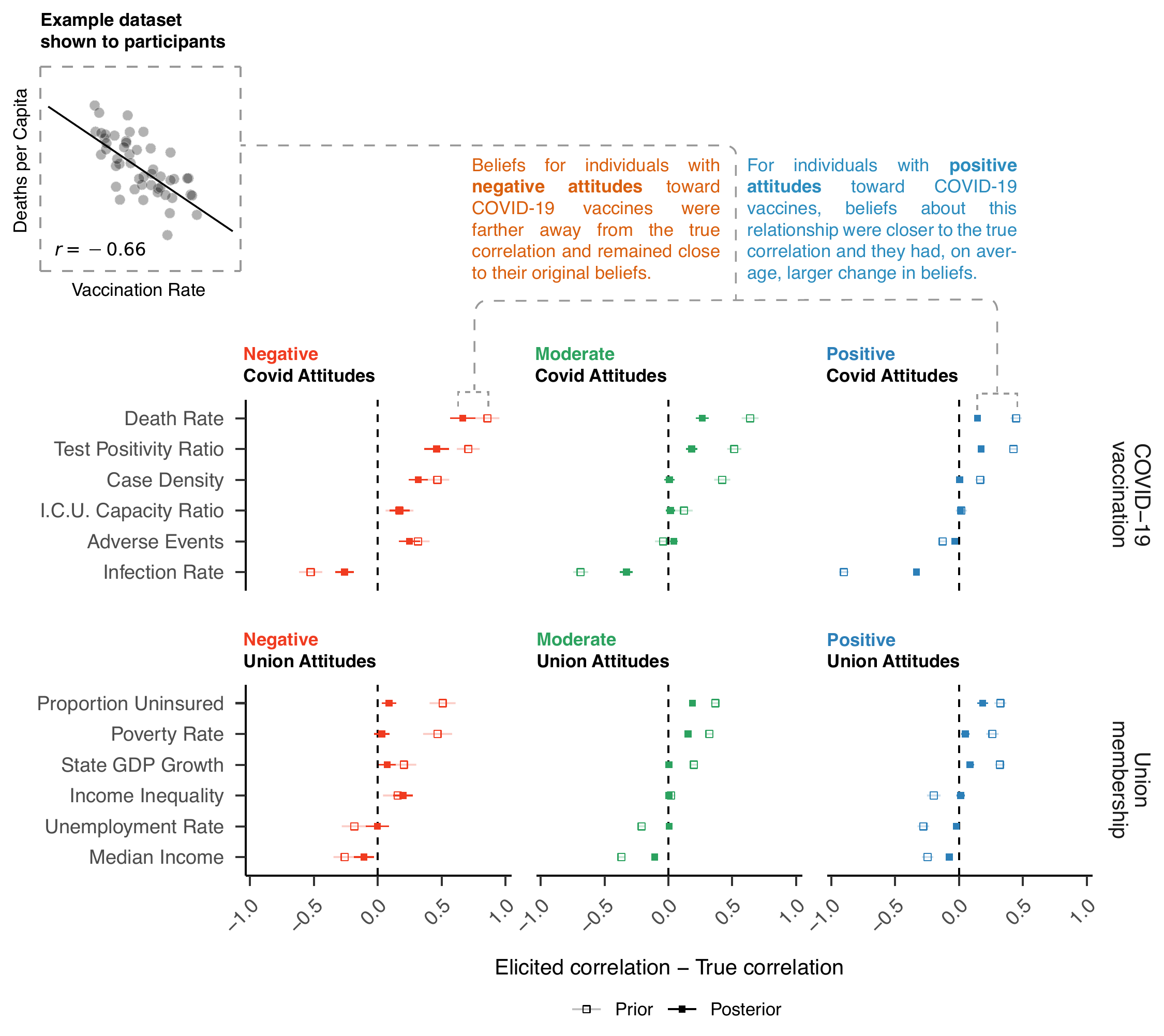}
\caption{Prior and posterior errors split by global attitude. \revised{Error bars indicate standard error of the mean.} Vertical black lines indicate no difference between elicited correlation and true correlation \revised{for each dataset listed at the left}.} 
\label{fig:beliefs_by_attitude}
\Description{The figure shows a comparison of prior and posterior beliefs, separated by participants' overall attitude toward each topic. This comparison shows that participants with strong anti-vaccination attitudes exhibited smaller adjustments in their beliefs about the COVID-19 variable sets.}

\end{figure*}

\subsubsection{Impact of prior attitude on prior and posterior belief errors}

For further insight into the effect of strong prior attitudes on belief change \revised{(RQ1)}, we examined the signed errors for both prior and posterior beliefs (i.e., the difference between the elicited correlation and the true correlation shown by the data) in Figure~\ref{fig:beliefs_by_attitude}, splitting participants into three groups according to their global attitude for each topic (negative, moderate, or positive, with the three categories spanning equally sized segments of the response scale).
Consistent with the results \revised{described} in the previous section, this indicated that the amount of belief change was influenced by the alignment of participants' attitudes with the provided evidence \revised{across several datasets for the COVID-19 vaccination topic}.
For example, consider the dataset for the relationship between COVID-19 vaccination rates and death rates.
This dataset depicts a strong negative correlation, which can be interpreted as evidence in favor of a positive attitude toward vaccination.
For this dataset, participants with negative preexisting attitudes exhibited more conservatism in belief updating compared to those with moderate and positive attitudes, with posterior beliefs that were close to their priors and which remained distant from the true correlation conveyed by the data.
This pattern was also apparent for the other COVID-19 datasets for Test Positivity, Case Density, \revised{I.C.U. Capacity}, and Adverse Vaccination Events.
\revised{Note that the Infection Rate dataset depicted a moderate positive correlation between vaccination rates and infection rates, which unlike the other datasets might be interpreted as evidence \emph{against} vaccination. This was the only dataset for the COVID-19 topic where participants with an anti-vaccination attitude had posterior beliefs that were at a similar distance to the true sample correlation compared to participants with moderate or positive attitudes.}

\revised{Taken together, these results suggest that participants with negative attitudes toward COVID-19 vaccination exhibited smaller changes in beliefs when viewing scatterplot visualizations that conflicted with their existing views. This resulted in posterior beliefs that were consistently} further from the true correlation compared to participants with moderate and positive attitudes. 
\revised{Notably, this pattern was far less apparent for the union membership topic. One possible exception is the Income Inequality dataset, which indicated no correlation between union membership rates and income inequality (i.e., a horizontal correlation line).
For this variable set, participants with an anti-union attitude continued to express belief in a weakly positive correlation between these variables. On the whole, however, attitudes about union membership (which, as we noted earlier, were weaker, less certain, and viewed as less personally important) appeared to have less of an impact on how people adjusted their beliefs about specific empirical relationships compared to the COVID-19 vaccination topic.}

\subsection{Changes in \textit{uncertainty} about the correlation \revised{(RQ1\&2)}}

\revised{The findings in the previous section indicate that topic-specific attitudes impacted how people updated their beliefs about the most likely relationship between a pair of variables, and that this effect was stronger for visualizations that included uncertainty representations. 
In this section we examine whether these effects also manifested in changes in \textit{uncertainty} about that relationship as expressed in the size of the confidence intervals measured in the prior and posterior elicitations.}

\begin{figure*}[t] 
\centering 
\includegraphics[width=1\textwidth]{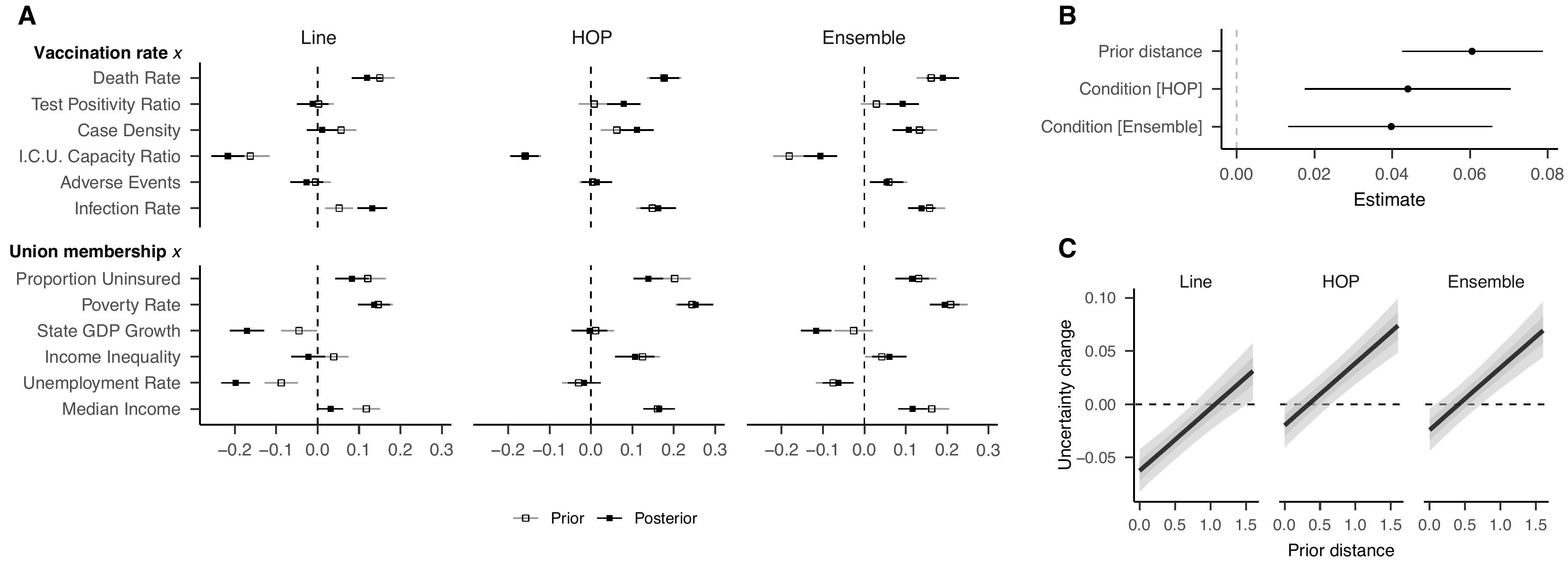}
\caption{\revised{\textbf{A:} Average changes in belief uncertainty (size of elicited CIs) for the prior and posterior elicitation, expressed as the difference between the elicited CI and the true CI for each dataset. Vertical lines indicates the elicited CI matches the size of the true CI. Error bars indicate standard error of the mean. \textbf{B:} Estimated coefficients for the uncertainty change model (posterior means and 95\% credible intervals). \textbf{C:} Expected predictive distributions from the uncertainty change model by visualization condition and prior distance.}} 
\label{fig:uncertainty_model}
\Description{The figure shows the results of the uncertainty change model (see full description in text).}
\end{figure*}

We conducted a similar analysis of changes in uncertainty from the prior to posterior beliefs as in Section~\ref{Belief_change} \revised{to explore the effects of topic-specific attitudes or visualization condition}.
We modeled the signed change in uncertainty ($CI^{post} - CI^{prior}$) using a t-distribution truncated at -2 and 2 with \textit{brms} \revised{(see Supplementary Materials Section \ref{supp:UC_models})}. 
Due to a technical error there were 88 trials (2\%) in which one or both of the CIs were not recorded and these trials were excluded from this analysis.

The best-performing model was the baseline model which included visualization condition and prior distance as fixed effects.
\revised{Figure~\ref{fig:uncertainty_model}B shows the estimated coefficients of this model, indicating a credible positive effect of prior distance on uncertainty change, such that uncertainty increased from prior to posterior elicitation when the data strongly conflicted with the prior belief.} 
Notably, the model comparison showed that there was no advantage to including attitude strength in the model (Supplementary Table S6).

The model also indicated credible effects of both uncertainty visualization conditions:
Compared to the Line condition, there were positive effects of both the HOP and Ensemble conditions on uncertainty changes (with no credible difference between those conditions). 
These effects are summarized in the posterior predictive distributions in Figure~\ref{fig:uncertainty_model}C.
\revised{In the Line condition where there was no visual representation of uncertainty, elicited uncertainty tended to decrease from prior to posterior responses, especially when the sample correlation was close to an individual's prior belief (i.e., prior distance close to zero).}
\revised{In contrast, visualizations with uncertainty representations were associated with} increases in subjective uncertainty about the relationship between variables, particularly when prior beliefs were different from the sample correlations depicted by the data.
\revised{However, it should be noted that} these effects were small relative to the size of the elicited CIs (median CI of .40 and .38 for prior and posterior elicitations, respectively).

Figure~\ref{fig:uncertainty_model}A shows the \revised{difference between the true CI associated with each dataset and} the prior and posterior elicited CIs for each variable pair. \revised{Positive values indicate cases where elicited CIs were larger compared to the true CI, while negative values indicate cases where people expressed less uncertainty compared to the true CI.}
In contrast to the results for beliefs about the most likely correlation (\revised{where responses in the posterior elicitation converged toward the sample correlation, see Figure~\ref{fig:beliefs_by_attitude}), prior and posterior CIs were relatively stable and of a similar magnitude across variable pairs.}
\revised{Surprisingly, even in the HOP and Ensemble conditions where the visualizations included uncertainty representations,} we did not observe that elicited CIs converged toward the true CIs for each dataset.
This suggests there was considerable variability across participants in how they expressed their uncertainty about the relationship \revised{even when they had direct access to} the uncertainty representations in the HOP and Ensemble conditions.

\subsection{Attitude change \revised{(RQ3)}}

Finally, we examined whether there were changes in attitudes from before
and after interacting with data visualizations for each topic
\textcolor{black}{(RQ3)}. There was no change in global attitude for
either COVID-19 vaccination
(\textcolor{black}{mean difference $D$ = -0.002 [-0.01, 0.01], $BF$ = 0.002})
or union membership
(\textcolor{black}{mean difference $D$ = 0.032 [-0.01, 0.07], $BF$ = 0.025}).
For attitude certainty there was no change for union membership
(\textcolor{black}{mean difference $D$ = -0.035 [-0.12, 0.05], $BF$ = 0.022}),
or for COVID-19 vaccination
(\textcolor{black}{mean difference $D$ = -0.002 [-0.01, 0.01], $BF$ = 0.002}).

Although there were no overall changes in attitudes toward union
membership, compared to the COVID-19 topic there was greater variability
in the pre/post changes in global attitudes
(\textcolor{black}{$\beta$ = 3.49 [2.69, 4.51], $BF$ = \ensuremath{8.849\times 10^{17}}})
and attitude certainty
(\textcolor{black}{$\beta$ = 2.37 [1.9, 3.04], $BF$ = \ensuremath{1.506\times 10^{12}}})
for the union topic. This suggests that attitudes about union membership
were more labile, consistent with the higher initial uncertainty and
lower topic involvement for that topic, even though the presented data
did not systematically lead to either more positive or negative
attitudes.



\section{Discussion}

Data visualizations are increasingly important tools for communicating science on a range of issues of pressing social concern.
Yet our understanding of their persuasive power remains limited, especially for audiences with closely held beliefs and attitudes that conflict with the intended message.
One reason for this is that it has been rare to measure changes in beliefs or attitudes that result from interactions with visualizations.
We built on prior work on belief and attitude elicitation \revised{\cite{karduni_bayesian_2021,karduni2023images,Pandey_TVCG_2014}} to measure individuals' global attitudes and their beliefs about \revised{specific statistical} relationships \revised{related to} two politically polarized topics: COVID-19 vaccination and labor union membership. 
\revised{We leveraged a graphical belief elicitation method to capture changes in beliefs about bivariate correlations} 
\cite{karduni_bayesian_2021} that resulted from viewing scatterplots, \revised{a simple, widespread form of statistical visualization that is also common in persuasive contexts such as social media and news articles.}

Our first research question focused on how existing attitudes about each topic affected how people updated their beliefs about correlations \revised{communicated} by the data.
\revised{Our attitude measures revealed important differences in individuals' existing views toward the two topics:} Compared to labor union membership, attitudes toward COVID-19 vaccination were more extreme and associated with a stronger sense of \revised{certainty} (perceived attitude clarity and correctness, see ~\cite{Petrocelli2007}) and increased personal relevance. \revised{Our main finding is that}
these differences in attitudes also related to whether people updated their beliefs about specific empirical relationships concerning each topic, as stronger attitudes were associated with less belief change after viewing visualizations that differed from prior beliefs (Figure~\ref{fig:belief_models}).
Further examination of these belief changes suggest that this effect was especially pronounced among participants who had negative attitudes toward COVID-19 vaccination (Figure~\ref{fig:beliefs_by_attitude}).
Those participants made smaller adjustments to their beliefs \revised{after observing} evidence that was favorable to COVID-19 vaccination, while participants with moderate or positive attitudes showed greater belief change and had posterior beliefs that were closer to the true correlations shown by the data.
\revised{This finding adds to recent evidence of conservatism in belief updating when the data presented in a scatterplot conflicts with an individuals' prior belief about a correlation~\cite{karduni_bayesian_2021,Xiong2022}.
The present work extends those efforts by showing that disparities between the data and prior beliefs alone cannot account for whether people exhibit conservatism in belief updating.
Overall attitudes on a topic---which may be rooted in aspects of social identity and worldview---may broadly influence whether people will change their beliefs about statistical relationships that are related to the same issue.
As such, the persuasiveness of statistical visualizations such as scatterplots will depend not only on the strength and clarity of the evidence portrayed by the data, but also the viewers' background and broader views about the topic.
}


Our second research question concerned the role of uncertainty representations in belief change.
Researchers have debated the merits of including representations of uncertainty (e.g., confidence intervals) in visualization~\cite{Hullman8805422} and other forms of science communication~\cite{daoust_should_2021,kelp_vaccinate_2022}.
We compared a baseline Line condition \revised{(with a scatterplot of data points and superimposed correlation line)} against conditions where the statistical uncertainty was represented in either a static distribution of lines (Ensemble condition) or an animated Hypothetical Outcome Plot (HOP condition). 
We found support for our hypothesis that uncertainty representations would lead to less belief change when prior beliefs differed from the data: \revised{Our statistical model of absolute belief change indicated credible interactions between} attitude strength and visualization condition, such that the effect of attitude strength was stronger in the Ensemble and HOP conditions compared to the the Line condition (Figure~\ref{fig:belief_models}).
In other words, uncertainty representations had no effect when the data conflicted with prior beliefs but the viewer did not have a strong preexisting attitude about the topic.
It was only when people were confronted with data that conflicted with their prior belief about the relationship \textit{and} they had a strong attitude that uncertainty representations led to smaller belief changes.
\revised{A potential explanation for this effect is that people with strong attitudes are more likely to attend to the uncertainty representation (e.g., examining the distribution of alternative correlation lines in the Ensemble condition) because they are motivated to find support for their existing belief about the correlation, leading them to assign less weight to the most likely correlation when updating their beliefs (for a related effect of Ensemble visualizations see~\cite{padilla2017effects}).
In contrast, individuals with weaker attitudes, who tended to have less confidence in their prior belief (Figure~\ref{fig:attitudes}B), may have focused more on the correlation line representing the most likely relationship while being less motivated to inspect the uncertainty representation.}

We also found evidence that uncertainty visualizations had a positive effect on uncertainty about the correlations compared to the Line condition (Figure~\ref{fig:uncertainty_model}, right).
In the Ensemble and HOP conditions uncertainty increased when prior beliefs were distant from the true correlation, while in the Line condition uncertainty decreased when the data reinforced prior beliefs.
However, in contrast to changes in beliefs about the most likely correlation, changes in uncertainty were unrelated to participants' attitudes and were relatively stable from prior to posterior elicitation, without the same convergence toward the true values as seen for the most likely correlation.
This is especially notable for the Ensemble condition in which participants were provided a static representation of the actual CI for a given dataset, yet participants continued to express greater uncertainty about the correlation.
This implies that participants were not simply matching their response in the posterior elicitation to the displayed CI, but further work is needed to understand how people make use of this belief elicitation technique to express their uncertainty.

Finally, we explored whether attitudes would shift as a result of interacting with the datasets\revised{, but found no evidence for changes in attitude valence or certainty for either topic.}
While there were no systematic changes in overall attitudes toward Union membership, attitudes were more variable for that topic compared to the COVID-19 topic, \revised{which might indicate that participants' views about union membership were more malleable} and influenced by their interactions with the data.
There may be several \revised{reasons} that we didn't see clearer effects on attitudes.
\revised{While our COVID-19 datasets presented a largely favorable view of COVID-19 vaccination, a large number of our participants already had strongly pro-vaccination views, leaving less opportunity to observe attitude changes for the topic.}
\revised{In addition,} our primary goal in designing the study was to present participants with trustworthy data that would ``speak for itself.'' 
To that end we used real-world datasets and provided source information to enhance the credibility of the visualizations, \revised{but we did not add any} further persuasive messaging \revised{to reinforce a positive or negative interpretation of the data}.
\revised{Lastly, we did not ask participants to actively make judgments or interact with the visualizations. Research on persuasive communication suggests that elaborative processing of a message, in which the receiver actively generates interpretations of the message content or connects it to their existing knowledge, is an important driver of attitude change~\cite{petty1986communication}, and that} brief interventions targeting specific beliefs are often unsuccessful in changing well-established attitudes in the absence of such elaborative processing~\cite{Albarracin:2005}.
Thus, the present findings highlight the need for further research on \revised{how principles of persuasive communication can be used to design} visualizations that both inform and create lasting changes in broader attitudes.

\subsection{Limitations}
An important limitation in our study is that even though beliefs about correlations can be \revised{well-defined} as values between -1 to 1, it is much more difficult to conceptualize attitudes as such bounded continuous values. Since each individual is self-reporting their own attitudes, comparison between them becomes noisier. For example, consider a person self-reporting an attitude toward COVID-19 vaccines representing the maximum negative value in our bounded scale. It is entirely possible that other individuals might have much more negative prior attitudes. Moreover, it is possible the stronger the conviction and prior attitude, the more pronounced the impact on belief change. Better approaches for measuring the intensity and multidimensional structure of topic-specific attitudes is an important area of improvement for future work.

One alternative interpretation of our results is that in the posterior belief elicitation people are not reporting changes in their subjective beliefs, but instead are simply attempting to \revised{recall and reproduce} the sample correlation they just observed.
Although we instructed participants to report their genuine beliefs about the correlation, it can't be ruled out that some participants' posterior beliefs were anchored to the \revised{sample correlation} shown with the data.
\revised{One way to rectify this issue would be to alter the task such that participants were incentivized to express their belief about the relationship rather than match the observed dataset (e.g., by being rewarded for accurate predictions about other datasets involving the same variables).}
\revised{A related concern is that any apparent belief changes observed in this task are short-lived and dependent on participants' ability to remember the scatterplot visualization. An important goal of future work is to measure} whether changes in beliefs persist over time and \revised{how any such changes depend on memory of the visualizations}.


\subsection{Implications and Future Directions}

Our results highlight that the way we consume and make inferences from data visualizations is not separable from the social and political constructs we inhabit.
\revised{As with other sources of evidence, whether a visualization is aligned with a viewer's attitudes and motivations is likely to be a crucial determinant of its persuasive power.}  
This point is further complicated by the potential interaction between uncertainty representations and belief change. In her work on why authors do not visualize uncertainty, Hullman \cite{Hullman8805422} presented several reasons visualization why authors might not include uncertainty representations in their designs. Specifically, Tenet 3 - "Rhetorical Model of Uncertainty Omission" discusses a related point: The perception that uncertainty representations might cast doubt on the "signal" from visualizations is a primary factor in visualization designers' choice not to make use of them. 
One might argue that our results offer support to such perceptions. However, it is important to point out that it was only under strong prior attitudes that uncertainty representations were associated with reduced belief change. 
Attitudes, on the other hand, are notoriously hard to change and are influenced by factors such as moral conviction and political identity \cite{skitka2015psychological} that are potentially unrelated to the specific visualizations. 
Thus, we argue that instead of not including uncertainty representations, 
we should aim to understand the psychological and social barriers to attitude change and their implications for how people interact with and evaluate data visualizations.

The intersection of data visualization and persuasion is an important area for further research. Recent research on reasoning with misinformation suggests that deliberation and elaboration might shift our attention to more accurate information \cite{pennycookLazyNotBiased2019, pennycook2021shifting}. This suggests that there could be stronger persuasive effects if the data visualizations are combined with persuasive messages or interactions that encourage elaborative thinking. In the future, we are interested in investigating whether encouraging users to reason about or explain the data, eliciting beliefs and attitudes, and including persuasive messages are potential interventions that can increase elaborative processing of data visualizations and lead to attitude change.

\section{Conclusion}

We investigated the impacts of prior attitudes and uncertainty representations on the persuasive power of visualization in terms of belief and attitude change. 
We found that strong prior attitudes and uncertainty representations lead to smaller belief changes in light of incongruent data. 
Taken together, our results suggest that participants with stronger prior attitudes are less likely to update their beliefs \revised{about statistical relationships} when presented with \revised{scatterplot} visualizations, \revised{and that this effect is magnified when those visualizations include uncertainty representations}. 
\revised{We also found little evidence of changes in attitudes as a result of viewing multiple related datasets on a topic.}
\revised{Given the importance of visualizations in science communication, more research is needed to understand how visualizations prompt changes in both specific beliefs and broader attitudes.} Our results highlight the importance of situating visualization research and designs within the larger social, political, and ethical context of persuasion.

\bibliographystyle{ACM-Reference-Format}

\end{document}


\title{Supplementary Materials for: When do data visualizations persuade? The impact of prior attitudes on \revised{learning about correlations} from \revised{scatterplot} visualizations}


\author{Douglas Markant}
\email{dmarkant@uncc.edu}
\orcid{0000-0003-0568-2648}
\authornotemark[1]
\affiliation{%
  \institution{University of North Carolina at Charlotte}
}

\author{Milad Rogha}
\email{mrogha@uncc.edu}
\orcid{0000-0002-1464-2157}
\affiliation{%
  \institution{University of North Carolina at Charlotte}
}

\author{Alireza Karduni}
\email{akarduni@uncc.edu}
\orcid{0000-0001-9719-7513}
\affiliation{%
  \institution{IDEO}
}

\author{Ryan Wesslen}
\email{ryan@explosion.ai}
\orcid{0000-0001-9638-8078}
\affiliation{%
 \institution{Explosion}
}

\author{Wenwen Dou}
\email{wdou1@uncc.edu}
\orcid{0000-0003-0319-9484}
\affiliation{%
  \institution{University of North Carolina at Charlotte}
  }




\maketitle


\FloatBarrier
\newpage

\renewcommand{\thefigure}{S\arabic{figure}}
\setcounter{figure}{0}

\renewcommand{\thetable}{S\arabic{table}}
\setcounter{table}{0}

\section{Supplementary Materials}

\begin{figure}[h]
\centering
\includegraphics[width=.6\textwidth]{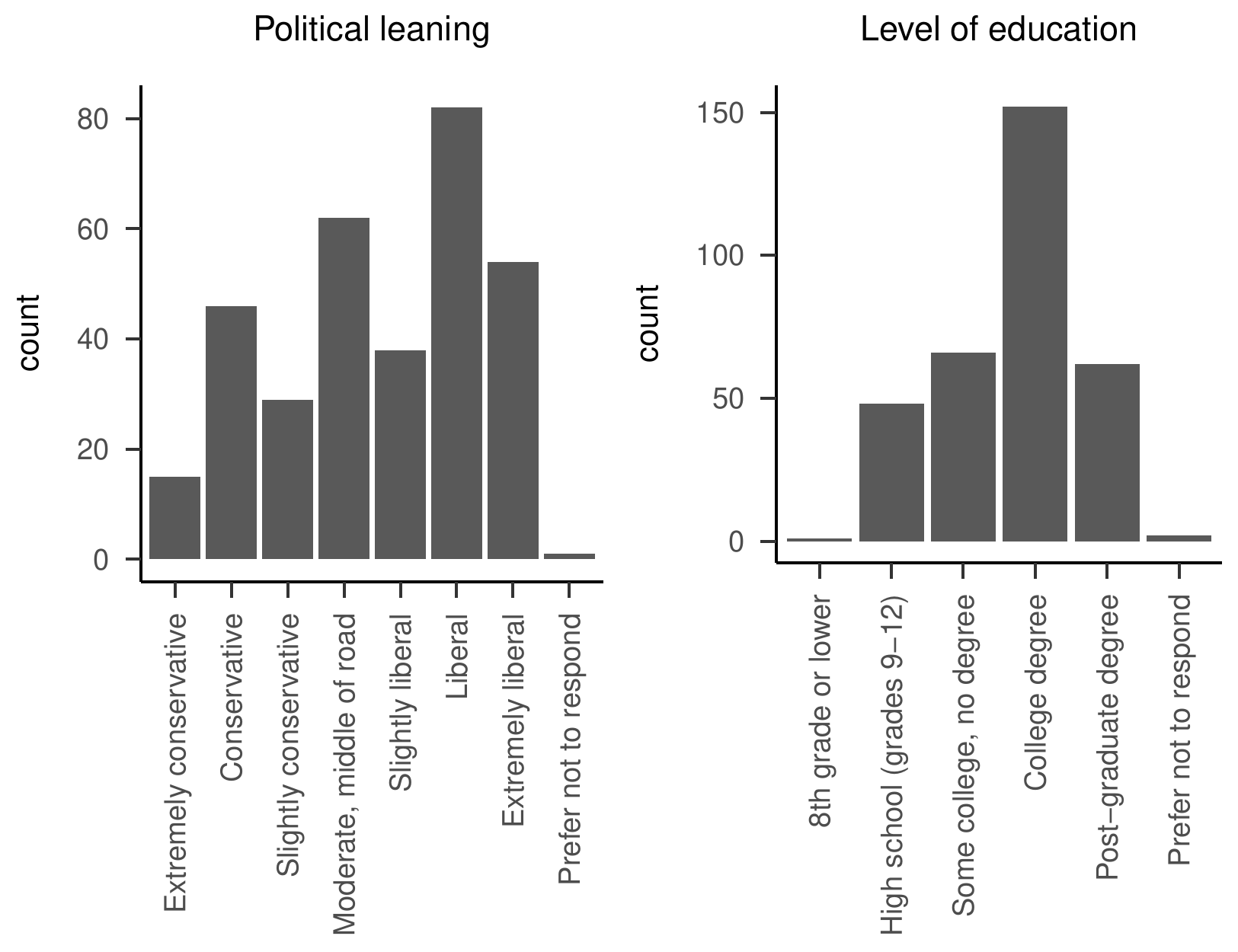}
\caption{Histogram of sample demographics for political leaning and level of education.}
\label{fig:demographics}
\end{figure}

\begin{table}[tbp]

\begin{center}
\begin{threeparttable}

\caption{Descriptive statistics for pre/post attitudes and involvement.}

\begin{tabular}{lllll}
\toprule
 & \multicolumn{2}{c}{Pre} & \multicolumn{2}{c}{Post} \\
\cmidrule(r){2-3} \cmidrule(r){4-5}
 & Mean & SD & Mean & SD\\
\midrule
Attitude (Union) & 1.26 & 2.10 & 1.30 & 2.02\\
Clarity (Union) & 6.63 & 1.95 & 6.55 & 1.79\\
Correctness (Union) & 5.85 & 1.95 & 5.89 & 1.93\\
Topic involvement (Union) & 12.54 & 4.20 & 12.91 & 4.26\\
Attitude (COVID-19) & 2.24 & 2.52 & 2.26 & 2.47\\
Clarity (COVID-19) & 8.15 & 1.13 & 7.95 & 1.28\\
Correctness (COVID-19) & 7.32 & 1.54 & 7.27 & 1.66\\
Topic involvement (COVID-19) & 15.39 & 3.58 & 15.60 & 3.60\\
\bottomrule
\end{tabular}

\end{threeparttable}
\end{center}

\end{table}


\clearpage

\subsection{Relationships between topic-specific attitudes and individual characteristics}
\label{supp:covariates}

\documentclass[
]{article}
\usepackage{amsmath,amssymb}
\usepackage{lmodern}
\usepackage{iftex}
\ifPDFTeX
  \usepackage[T1]{fontenc}
  \usepackage[utf8]{inputenc}
  \usepackage{textcomp} 
\else 
  \usepackage{unicode-math}
  \defaultfontfeatures{Scale=MatchLowercase}
  \defaultfontfeatures[\rmfamily]{Ligatures=TeX,Scale=1}
\fi
\IfFileExists{upquote.sty}{\usepackage{upquote}}{}
\IfFileExists{microtype.sty}{
  \usepackage[]{microtype}
  \UseMicrotypeSet[protrusion]{basicmath} 
}{}
\makeatletter
\@ifundefined{KOMAClassName}{
  \IfFileExists{parskip.sty}{%
    \usepackage{parskip}
  }{
    \setlength{\parindent}{0pt}
    \setlength{\parskip}{6pt plus 2pt minus 1pt}}
}{
  \KOMAoptions{parskip=half}}
\makeatother
\usepackage{xcolor}
\usepackage[margin=1in]{geometry}
\usepackage{graphicx}
\makeatletter
\def\maxwidth{\ifdim\Gin@nat@width>\linewidth\linewidth\else\Gin@nat@width\fi}
\def\maxheight{\ifdim\Gin@nat@height>\textheight\textheight\else\Gin@nat@height\fi}
\makeatother
\setkeys{Gin}{width=\maxwidth,height=\maxheight,keepaspectratio}
\makeatletter
\def\fps@figure{htbp}
\makeatother
\setlength{\emergencystretch}{3em} 
\providecommand{\tightlist}{%
  \setlength{\itemsep}{0pt}\setlength{\parskip}{0pt}}
\setcounter{secnumdepth}{-\maxdimen} 
\usepackage{color}
\usepackage{soul}
\ifLuaTeX
  \usepackage{selnolig}  
\fi
\IfFileExists{bookmark.sty}{\usepackage{bookmark}}{\usepackage{hyperref}}
\IfFileExists{xurl.sty}{\usepackage{xurl}}{} 
\urlstyle{same} 
\hypersetup{
  hidelinks,
  pdfcreator={LaTeX via pandoc}}

\author{}
\date{\vspace{-2.5em}}

\begin{document}

Descriptive statistics for participant characteristics (political
leaning, epistemic beliefs, defensive confidence, and CRT) can be found
in Table\textasciitilde{}\ref{tab:desc.ind}.

Global attitudes were correlated with a number of individual
characteristics. Liberal political orientation was associated with more
positive attitudes toward both union membership
\textcolor{black}{($r$ = 0.38 [0.27, 0.48], $BF$ = \ensuremath{1.64\times 10^{8}})}
and COVID-19 vaccination
\textcolor{black}{($r$ = 0.23 [0.12, 0.34], $BF$ = 220.31)}. Attitude
valence was also associated with different aspects of epistemic beliefs.
Believing that ``truth is political'' was associated with more negative
attitudes toward both union membership
\textcolor{black}{($r$ = -0.19 [-0.3, -0.09], $BF$ = 57.84)} and COVID-19
vaccination \textcolor{black}{($r$ = -0.18 [-0.29, -0.08], $BF$ = 36.42)},
while endorsing the ``need for evidence'' was associated with more
positive attitudes toward COVID-19 vaccination
\textcolor{black}{($r$ = 0.23 [0.13, 0.33], $BF$ = 1709.94)} and union
membership \textcolor{black}{($r$ = 0.25 [0.15, 0.35], $BF$ = 7064.1)}.
``Faith in intuition'' was not associated with attitudes toward COVID-19
vaccination \textcolor{black}{($r$ = -0.09 [-0.2, 0.02], $BF$ = 0.44)} or
union membership
\textcolor{black}{($r$ = -0.09 [-0.2, 0.01], $BF$ = 0.47)}.

People who expressed greater defensive confidence also tended to have
higher attitude clarity (Union:
\textcolor{black}{$r$ = 0.34 [0.24, 0.43], $BF$ = \ensuremath{1.23\times 10^{8}}}
; COVID-19:
\textcolor{black}{$r$ = 0.33 [0.24, 0.43], $BF$ = \ensuremath{3.92\times 10^{7}}}
) and higher attitude correctness (Union:
\textcolor{black}{$r$ = 0.21 [0.11, 0.31], $BF$ = 219.21} ; COVID-19:
\textcolor{black}{$r$ = 0.19 [0.08, 0.29], $BF$ = 78.04} ).
Lastly, we did not find evidence that attitudes
were associated with intuitive vs.~analytic thinking as assessed by the
CRT (all \(BF < 1\)).

\end{document}

\begin{table}[tbp]
\begin{center}
\begin{threeparttable}

\caption{Descriptive statistics for individual measures}
\label{tab:desc.ind}

\begin{tabular}{lllll}
\toprule
 & \multicolumn{1}{c}{Mean} & \multicolumn{1}{c}{SD} & \multicolumn{1}{c}{Min} & \multicolumn{1}{c}{Max}\\
\midrule
Political affiliation (-3 = Extremely conservative, 3 = Extremely liberal) & 0.61 & 1.82 & -3.00 & 3.00\\
Defensive confidence & 45.35 & 7.98 & 23.00 & 60.00\\
EB - Faith in intuition & 3.21 & 0.81 & 1.00 & 5.00\\
EB - Need for evidence & 4.21 & 0.74 & 1.00 & 5.00\\
EB - Truth is political & 2.47 & 0.99 & 1.00 & 5.00\\
CRT - Correct responses & 2.56 & 1.07 & 0.00 & 4.00\\
CRT - Intuitive responses & 1.15 & 1.04 & 0.00 & 4.00\\
\bottomrule
\addlinespace
\end{tabular}

\begin{tablenotes}[para]
\normalsize{\textit{Note.} EB: Epistemic beliefs; CRT: Cognitive Reflection Test}
\end{tablenotes}

\end{threeparttable}
\end{center}

\end{table}

\clearpage

\subsection{Absolute belief change models}
\label{ABC_model_supplement}

Absolute belief changes ranged from 0 to 2 and were transformed to the 0--1 interval in order to perform Beta regression \revised{(\(|r^{post} - r^{prior}|/2\))}. 
After transformation the data included 44 trials (1.5\%) in which belief change was equal to 0 or 1, which lie outside the range modeled by Beta regression.
We therefore performed zero-one-inflated Beta (ZOIB) regression which includes additional parameters for the probability of a 0/1 response ($\alpha$) and, among those extreme responses, the probability of 1 ($\gamma$). 
Given the small number of cases we simply modeled these parameters with intercepts ($\alpha \sim 1$ and $\gamma \sim 1$).

The Beta component of the model includes two parameters for the mean ($\mu$) and precision ($\phi$) of the Beta distribution.
\revised{The mean was modeled with fixed effects for the factors identified in our preregistration: the absolute sample correlation of the dataset ($SC$),} visualization condition ($VC$: Line, HOP, Ensemble), the difference between the elicited prior belief and sample correlation (prior distance, $PD$), and attitude strength ($AS$).
\revised{However, in our preregistration we did not enumerate the full set of models that we compared here, which we selected to cover the space of possible interactions between visualization condition, prior distance, and attitude strength.}
The fixed effects structures for the full set of models are shown in Table S3.
All models also included random effects for participant ID ($\mu \sim ... + (1|pid)$).

Preliminary modeling with only an intercept for the precision parameter ($\phi \sim 1$) indicated that the model failed to account for a narrow peak close to zero (as assessed by posterior predictive checks).
\revised{We therefore allowed the precision of the distribution to vary as a function of prior distance ($\phi \sim PD$) based on the reasoning that there would naturally be less variability in the extent of belief changes when prior beliefs were already close to the true sample correlation.}

\begin{table}[H]

\begin{center}
\begin{threeparttable}

\caption{Fixed effects formulas for absolute belief change models.}

\begin{tabular}{ll}
\toprule
 \multicolumn{1}{c}{Model} & \multicolumn{1}{c}{Formula} \\
\midrule

ABC-0     & \[\mu \sim SC + VC + PD\] \\
ABC-1     & \[\mu \sim SC + VC + PD + VC \times PD\] \\ 
ABC-2     & \[\mu \sim SC + VC + PD + AS\] \\ 
ABC-3     & \[\mu \sim SC + VC + PD + AS + VC \times PD\] \\ 
ABC-4     & \[\mu \sim SC + VC + PD + AS + VC \times AS\] \\ 
ABC-5     & \[\mu \sim SC + VC + PD + AS + PD \times AS\] \\ 
ABC-6     & \[\mu \sim SC + VC + PD + AS + VC \times PD + VC \times AS\] \\ 
ABC-7     & \[\mu \sim SC + VC + PD + AS + VC \times PD + PD \times AS\] \\ 
ABC-8     & \[\mu \sim SC + VC + PD + AS + VC \times AS + PD \times AS\] \\ 
ABC-9     & \[\mu \sim SC + VC + PD + AS + VC \times PD + VC \times AS + PD \times AS\] \\ 
ABC-10    & \[\mu \sim SC + VC + PD + AS + VC \times PD + VC \times AS + PD \times AS + VC \times PD \times AS\] \\ 

\bottomrule
\end{tabular}

\label{tab:belief_change_model_formulas}
\end{threeparttable}
\end{center}

\end{table}

\begin{table}[tbp]

\begin{center}
\begin{threeparttable}

\caption{Results of model comparison for absolute belief change.}

\begin{tabular}{lllll}
\toprule
 & \multicolumn{1}{c}{looic} & \multicolumn{1}{c}{elpd\_loo} & \multicolumn{1}{c}{elpd\_diff} & \multicolumn{1}{c}{se\_diff}\\
\midrule
ABC-10 & -6,452.85 & 3,226.42 & 0.00 & 0.00\\
ABC-7 & -6,450.70 & 3,225.35 & -1.07 & 3.23\\
ABC-5 & -6,450.70 & 3,225.35 & -1.08 & 3.65\\
ABC-9 & -6,448.13 & 3,224.07 & -2.36 & 3.16\\
ABC-8 & -6,447.36 & 3,223.68 & -2.74 & 3.61\\
ABC-2 & -6,442.67 & 3,221.33 & -5.09 & 5.64\\
ABC-3 & -6,441.70 & 3,220.85 & -5.58 & 5.18\\
ABC-6 & -6,439.72 & 3,219.86 & -6.56 & 5.16\\
ABC-4 & -6,438.74 & 3,219.37 & -7.06 & 5.65\\
ABC-1 & -6,417.43 & 3,208.71 & -17.71 & 7.22\\
ABC-0 & -6,415.94 & 3,207.97 & -18.45 & 7.59\\
\bottomrule
\end{tabular}

\end{threeparttable}
\end{center}

\end{table}






\begin{table}[tbp]

\begin{center}
\begin{threeparttable}

\caption{Full set of estimated fixed effects for absolute belief change model (ABC-10).}
\label{tab:belief_change_model_coefficients}
\begin{tabular}{llll}
\toprule
 & \multicolumn{1}{c}{Estimate} & \multicolumn{1}{c}{l-95\% CI} & \multicolumn{1}{c}{u-95\% CI}\\
\midrule
Intercept & -2.36 & -2.55 & -2.16\\
Abs. sample correlation & -0.46 & -0.59 & -0.34\\
Prior distance & 2.28 & 1.98 & 2.59\\
Attitude strength & -0.08 & -0.14 & -0.01\\
Condition(HOP) & -0.12 & -0.39 & 0.14\\
Condition(Ensemble) & -0.32 & -0.58 & -0.06\\
Prior distance X Attitude strength & 0.01 & -0.08 & 0.11\\
Prior distance X Condition(HOP) & 0.28 & -0.15 & 0.72\\
Prior distance X Condition(Ensemble) & 0.33 & -0.08 & 0.74\\
Attitude strength X Condition(HOP) & 0.07 & -0.02 & 0.16\\
Attitude strength X Condition(Ensemble) & 0.11 & 0.02 & 0.19\\
Prior distance X Attitude strength X Condition(HOP) & -0.16 & -0.30 & -0.01\\
Prior distance X Attitude strength X Condition(Ensemble) & -0.18 & -0.32 & -0.05\\
phi - Intercept & 2.62 & 2.54 & 2.70\\
phi - Prior distance & -1.09 & -1.21 & -0.98\\
alpha - Intercept & -4.00 & -4.24 & -3.77\\
gamma - Intercept & -0.92 & -1.45 & -0.42\\
\bottomrule
\end{tabular}

\end{threeparttable}
\end{center}

\end{table}

\clearpage

\begin{figure}[t]
\centering
\includegraphics[width=1\textwidth]{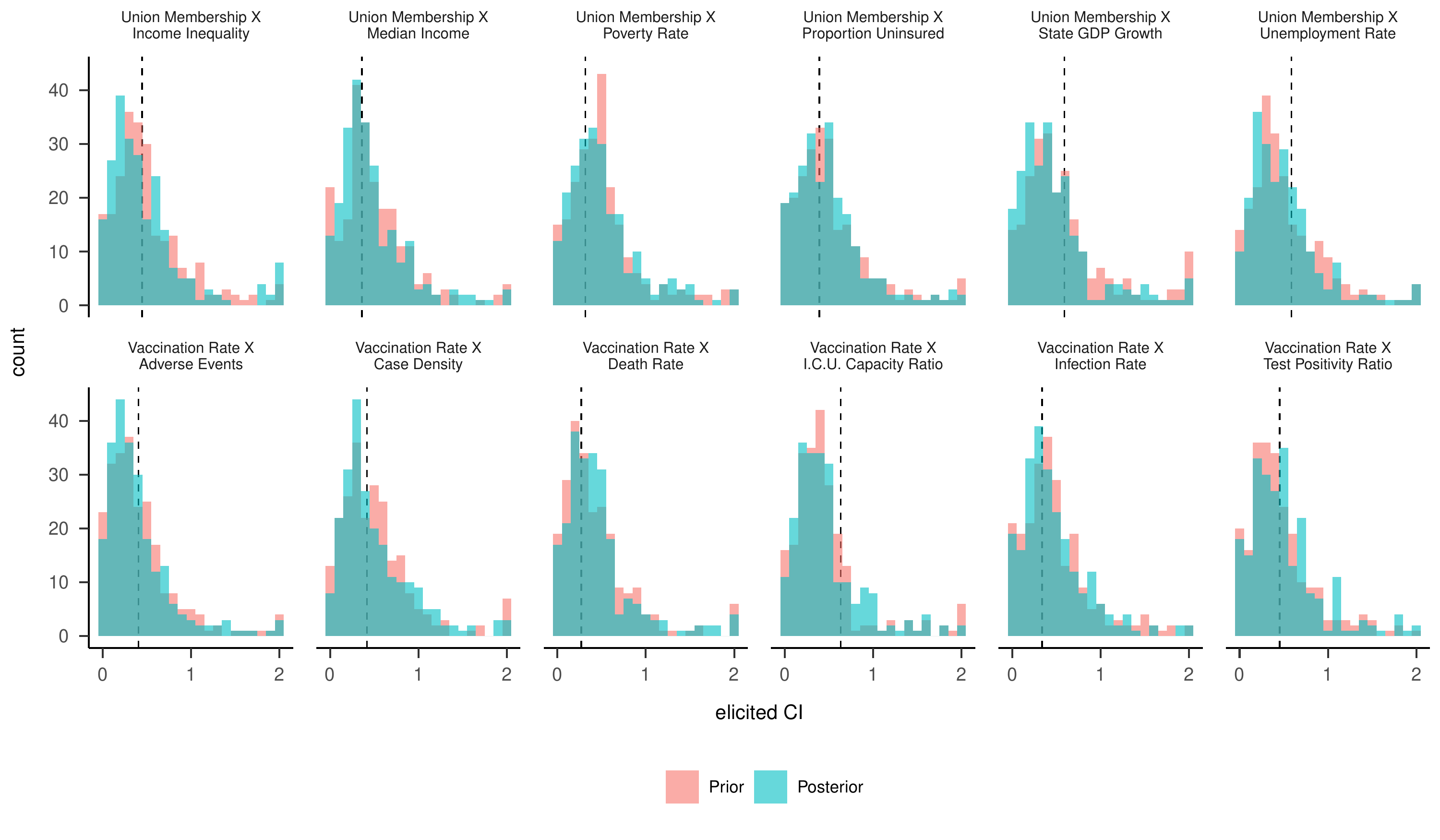}
\caption{Distributions of prior and posterior CIs for each variable pair.}
\label{fig:beliefs_uncertainty_prepost}
\end{figure}

\clearpage

\subsection{Uncertainty change models}
\label{supp:UC_models}

\revised{Uncertainty changes were calculated as the difference between the magnitude of the elicited CIs ($CI^{post} - CI^{prior}$) for each trial.
Uncertainty changes were modeled using a Student-t-distribution truncated at $-2$ and $+2$ with \textit{brms}.
The fixed effects formulas for the models included in the comparison are shown in Table S6, with predictors for visualization condition ($VC$), prior distance ($PD$), and attitude strength ($AS$).
All models additionally included random effects for participant ID ($\mu \sim ... + (1|pid)$).
Two of the models (UC-9 and UC-10) did not converge and are therefore not included in the results.} 

\begin{table}[h]
\begin{center}
\begin{threeparttable}
\caption{Fixed effects formulas for uncertainty change models.}

\begin{tabular}{ll}
\toprule
 \multicolumn{1}{c}{Model} & \multicolumn{1}{c}{Formula} \\
\midrule

UC-0     & \[\mu \sim VC + PD\] \\
UC-1     & \[\mu \sim VC + PD + VC \times PD\] \\ 
UC-2     & \[\mu \sim VC + PD + AS\] \\ 
UC-3     & \[\mu \sim VC + PD + AS + VC \times PD\] \\ 
UC-4     & \[\mu \sim VC + PD + AS + VC \times AS\] \\ 
UC-5     & \[\mu \sim VC + PD + AS + PD \times AS\] \\ 
UC-6     & \[\mu \sim VC + PD + AS + VC \times PD + VC \times AS\] \\ 
UC-7     & \[\mu \sim VC + PD + AS + VC \times PD + PD \times AS\] \\ 
UC-8     & \[\mu \sim VC + PD + AS + VC \times AS + PD \times AS\] \\ 
UC-9     & \[\mu \sim VC + PD + AS + VC \times PD + VC \times AS + PD \times AS\] \\ 
UC-10    & \[\mu \sim VC + PD + AS + VC \times PD + VC \times AS + PD \times AS + VC \times PD \times AS\] \\ 

\bottomrule
\end{tabular}

\label{tab:uncertainty_change_model_formulas}
\end{threeparttable}
\end{center}

\end{table}

\begin{table}[tbp]

\begin{center}
\begin{threeparttable}

\caption{Model comparison for belief uncertainty change.}

\begin{tabular}{lllll}
\toprule
 & \multicolumn{1}{c}{looic} & \multicolumn{1}{c}{elpd\_loo} & \multicolumn{1}{c}{elpd\_diff} & \multicolumn{1}{c}{se\_diff}\\
\midrule
UC-0 & 2,124.52 & -1,062.26 & 0.00 & 0.00\\
UC-1 & 2,125.04 & -1,062.52 & -0.26 & 2.09\\
UC-2 & 2,126.94 & -1,063.47 & -1.21 & 0.40\\
UC-3 & 2,126.98 & -1,063.49 & -1.23 & 2.11\\
UC-7 & 2,128.23 & -1,064.11 & -1.85 & 2.15\\
UC-5 & 2,129.31 & -1,064.66 & -2.40 & 0.45\\
UC-6 & 6,742.20 & -3,371.10 & -2,308.84 & 90.04\\
UC-4 & 15,832.72 & -7,916.36 & -6,854.10 & 156.94\\
UC-8 & 234,202.14 & -117,101.07 & -116,038.81 & 427.35\\
\bottomrule
\end{tabular}

\end{threeparttable}
\end{center}

\end{table}